\documentclass[useAMS,usenatbib,usegraphicx]{mn2e}

\usepackage{aas_macros}
\usepackage{amsfonts,amssymb}
\usepackage{amsmath}
\usepackage{url}

\newcommand{\mathsc}[1]{\text{\textsc{#1}}} 
\DeclareRobustCommand{\ion}[2]{%
\relax\ifmmode
\ifx\testbx\f 
{\mathbf{#1\,\mathsc{#2}}}\else
{\mathrm{#1\,\mathsc{#2}}}\fi
\else\textup{#1\,{\mdseries\textsc{#2}}}%
\fi}

\title[Enabling parallel computing in CRASH]{Enabling parallel computing in CRASH}
\author[A. M. Partl, A. Maselli, B. Ciardi, A. Ferrara, V. M\"uller]
      {A. M. Partl$^{1}$\thanks{E-mail: apartl@aip.de}, A. Maselli$^{2,5}$;\thanks{
       E-mail: antinulla@googlemail.com},
       B. Ciardi$^{3}$,
       A. Ferrara$^{4}$, and
       V. M\"uller$^{1}$\\
$^{1}$Astrophysikalisches Institut Potsdam, An der Sternwarte 16, Potsdam, 14482, Germany\\
$^{2}$Osservatorio Astrofisico di Arcetri, Largo Enrico Fermi 5, 50125, Firenze, Italy\\
$^{3}$Max-Planck-Institut f\"{u}r Astrophysik, Karl-Schwarzschild-Strasse 1, 85748 Garching, Germany\\
$^{4}$Scuola Normale Superiore, Piazza dei Cavalieri 7, 56126 Pisa, Italy\\
$^{5}$EVENT Lab for Neuroscience and Technology, Universitat de Barcelona, Passeig de la Vall d'Hebron 171, 08035 Barcelona, Spain\\
}
\begin{document}

\date{Accepted 2011 January 21. Received 2011 January 20; in original form 2010 November 5}

\pagerange{\pageref{firstpage}--\pageref{lastpage}} \pubyear{2009}

\maketitle

\label{firstpage}

\begin{abstract}
We present the new parallel version ({\tt pCRASH2}) of the cosmological radiative transfer code {\tt CRASH2} for distributed memory supercomputing facilities. The code is based on a static domain decomposition strategy inspired by geometric dilution of photons in the optical thin case that ensures a favourable performance speed-up with increasing number of computational cores. Linear speed-up is ensured as long as the number of radiation sources is equal to the number of computational cores or larger. The propagation of rays is segmented and rays are only propagated through one sub-domain per time step to guarantee an optimal balance between communication and computation. We have extensively checked {\tt pCRASH2} with a standardised set of test cases to validate the parallelisation scheme. The parallel version of {\tt CRASH2} can easily handle the propagation of radiation from a large number of sources and is ready for the extension of the ionisation network to species other than hydrogen and helium. 
\end{abstract}


\section{Introduction}

The field of computational cosmology has developed dramatically
during the last decades. Especially the evolution of the baryonic
physics has played an important role in the understanding of the
transition from the smooth early universe to the structured present one.
We are especially interested in the development of the intergalactic
ionising radiation field and the thermodynamic state of the baryonic
gas. First measurements of the epoch of reionisation are soon
expected from new radio interferometers such as LOFAR\footnote{http://www.lofar.org/}
or MWA\footnote{http://www.mwatelescope.org/}, which will become  
operative within one year.
The interpretation of these measurements requires, among others, a treatment
of radiative transfer coupled to cosmological structure formation.

Over the last decade, many different
numerical algorithms have emerged, allowing the continuum radiative transfer equation to be
solved for arbitrary geometries and density distributions. A substantial fraction of the
codes solve the radiative transfer (RT) equation on regular or adaptive grids (for example see \citet{Gnedin:2001fc,
Abel:2002ss, Razoumov:2002qc, Mellema:2006ve}).
Other codes developed schemes that introduce RT into the SPH formalism \citep{Pawlik:2008mw,
Petkova:2009wj, Hasegawa:2010bd} or into unstructured grids \citep{Ritzerveld:2003hi, Paardekooper:2010rp}
The large amount of different numerical strategies prompted a comparison of the
different methods on a standardised problem set. For results of this comparison project 
we refer the reader to \citet{Iliev:2006fk} and \citet{Iliev:2009yq}, where the performance of 11 cosmological 
RT and 10 radiation hydrodynamic codes are systematically studied.

Our straightforward and very flexible approach is based on a ray-tracing Monte Carlo (MC)
scheme. Exploiting the particle nature of a radiation field, it is possible to solve the RT equation
for arbitrary three-dimensional Cartesian grids and an arbitrary distribution of absorbers.
By describing the radiation field in terms of photons, which are then grouped into
photon packets containing a large number of photons each, it is possible to solve the RT
along one dimensional rays. With this strategy the explicit dependence on direction and position
can be avoided. Instead of directly solving for the intensity field 
only the interaction of photons with the gas contained in the cells needs to be modelled, as done in 
long and short characteristics algorithms \citep{Mellema:2006ve,Rijkhorst:2006bh,Whalen:2006ef}. 
MC ray-tracing schemes differ from short and long characteristic methods. Instead of casting rays
through the grid to each cell in the computational domain, the radiation field is described
statistically by shooting rays in random directions from the source. However unlike in fully
Monte Carlo transport schemes where the location of the photon matter interaction is determined by 
sampling the packet's mean free path, the packets are propagated and attenuated through the grid 
from cell to cell along rays until all the photons are absorbed or the packets exits the computational domain. 
This allows for an efficient
handling of multiple point sources and diffuse radiation fields, such as recombination radiation
or the ultraviolet (UV) background field. Additionally this statistical approach easily allows 
for sources with anisotropic radiation. A drawback of any Monte Carlo sampling method 
however is the introduction of numerical noise. By increasing the number of rays used
for the sampling of the radiation field though, numerical noise can be reduced at cost of
computational resources.

Such a ray-tracing MC scheme has been successfully implemented in our code {\tt CRASH2},
which is, to date, one of the main references among RT numerical methods used in cosmology.
{\tt CRASH} was first introduced by \citet{Ciardi:2001cr} to follow the evolution of hydrogen
ionisation for multiple sources under the assumption that hydrogen has a fixed temperature.
Then the code has been further developed by including the physics of helium chemistry,
temperature evolution, and background radiation \citep{Maselli:2003, Maselli:2005ph}. In its
latest version, {\tt CRASH2}, the numerical noise by the MC sampling has been greatly reduced
through the introduction of coloured photon packets \citep{Maselli:2009gd}.

The problems that are being solved with cosmological RT codes become larger and larger, in terms of computational cost.
Especially, the study of reionisation is a demanding task, since a vast number of sources 
and large volumes are needed to properly model the era of reionisation 
\citep{Baek:2009zr, McQuinn:2007ys, Trac:2007vn, Iliev:2006ud}.
Furthermore the addition of more and more physical processes to {\tt CRASH2} requires increased
precision in the solution. To study such computationally demanding
models with {\tt CRASH2}, the code needs support for parallel distributed memory computers. In this
paper we present the parallelisation strategy adopted for our MPI parallel version of the latest
version of the serial {\tt CRASH2} code, which we call {\tt pCRASH2}.

The paper is structured as follows. First we give a brief summary of the serial {\tt CRASH2} 
implementation in Section \ref{sec:CRASH2}. In Section \ref{sec:parStrategy} we review the
existing parallelisation strategies for MC ray-tracing codes and describe the approach taken
by {\tt pCRASH2}. In Section \ref{sec:performance} we extensively test the parallel implementation
against standardised test cases. We further study the scaling properties of the parallel code
in Section \ref{sec:scaling} and summarise our results in Section \ref{sec:conclusions}.
Throughout this paper we assume $h=0.7$.

\section{{\tt CRASH2}: Summary of the Algorithm}
\label{sec:CRASH2} 

In this Section we briefly summarise the {\tt CRASH2} code. A complete description of the
algorithm is found in  \citet{Maselli:2003} and in \citet{Maselli:2009gd}, with an additional detailed 
description of the implementation for the background radiation field given in \citet{Maselli:2005ph}.
We refer the interested reader to these papers for a full description of {\tt CRASH2}.

{\tt CRASH2} is a Monte-Carlo long-characteristics continuum RT code, which is based on
ray-tracing techniques on a three-dimensional Cartesian grid. Since many of the processes
involved in RT, like recombination emission or scattering processes, are probabilistic, Monte Carlo
methods are a straight forward choice in capturing these processes adequately. {\tt CRASH2} 
therefore relies heavily on the sampling of various probability distribution functions (PDFs) which describe several 
physical processes such as the distribution of photons from a source, reemission due to electron
recombination, and the emission of background field photons.
The numerical scheme follows the propagation of ionising radiation through an arbitrary H/He 
static density field and captures the evolution of the thermal and ionisation state of the gas on the
fly. The typical RT effects giving rise to spectral filtering, shadowing and self-shielding are 
naturally captured by the algorithm.

The radiation field is discretised into distinct energy packets, which can be seen as packets of 
photons. These photon packets are characterised by a propagation direction and their 
spectral energy content $E(\nu_j)$ as a function of discrete frequency bins $\nu_j$.
Both the radiation fields arising from multiple point sources, located arbitrarily in the box, and
from diffuse radiation fields such as the background field or radiation produced by recombining
electrons are discretised into such photon packets.

Each source emits photon packets according to its luminosity $L_s$ at regularly spaced time 
intervals $\Delta t$. The total energy radiated by one source during the total simulation time 
$t_{\mathrm{sim}}$ is $E_s = \int_{0}^{t_{\mathrm{sim}}} L_s (t_s) {\mathrm{d}} t_s$. For each
source, $E_s$ is distributed in $N_p$ photon packets. The energy emitted per source in one
time step is further distributed according to the source's spectral energy distribution function 
into $N_\nu$ frequency bins $\nu_j$. We call such a photon packet a coloured packet. Then
for each coloured packet produced by a source in one time step, an emission direction is 
determined according to the angular emission PDF of the source. Thus $N_p$ is the main 
control parameter in {\tt CRASH2} governing both the time resolution as well as the
spatial resolution of the radiation field. 

After a source produced a coloured packet, it is propagated through the given density field. Every 
time a coloured packet traverses a cell $k$, the length of the path within each crossed cell is calculated and 
the cell's optical depth to ionising continuum radiation
$\tau_{c}^{k}$ is determined by summing up the contribution of the different absorber
species (\ion{H}{i}, \ion{He}{i}, \ion{He}{ii}). The total number of photons absorbed in cell $k$ per
frequency bin $\nu_j$ is thus
\begin{equation}
\label{eq:NumPhotInCell}
N_{A,\gamma}^{(k)}=N_{T,\gamma}^{(k-1)}\left(\nu_{j}\right)\left[1-\mathrm{e}^{-\tau_{c}^{k}\left(\nu_{j}\right)}\right]
\end{equation}
where $N_{T,\gamma}^{(k-1)}$ is the number of photons transmitted through cell $k-1$. The total
number of absorbed photons is then distributed to the various species according to their contribution
to the cell's total optical depth. Before the packet is propagated to the next cell, the cell's ionisation
fractions and temperature are updated by solving the ionisation network for 
$\Delta x_{\mathrm{\ion{H}{i}}}$, $\Delta x_{\mathrm{\ion{He}{i}}}$, $\Delta x_{\mathrm{\ion{He}{ii}}}$,
and by solving for changes in the cells temperature $\Delta T$ due to photo-heating and the
changes in the number of free particles of the plasma. The number of recombining electrons 
$N_{\mathrm{rec}}$ is recorded as well and is used for the production of the 
diffuse recombination radiation. In addition to the discrete process of photoionisation, {\tt CRASH2} 
includes various continuous ionisation and cooling processes in the ionisation network 
(bremsstrahlung, Compton cooling/heating, collisional ionisation, collisional ionisation cooling, 
collisional excitation cooling, and recombination cooling).

After these steps, the photon packet is propagated to the next cell and these steps are
repeated until the packet is either extinguished or, if periodic boundary conditions are not considered,
until it leaves the simulation box. At fixed time intervals $\Delta t_{\mathrm{rec}}$, the grid is checked
for any cell that has experienced enough recombination events to reach a certain threshold criteria
$N_{\mathrm{rec}} \ge f_{\mathrm{rec}} N_a$, where $N_a$ is the total number of species "$a$" atoms
and $f_{\mathrm{rec}} \in [0,1]$ is the recombination threshold. If the reemission criteria is
fulfilled, a recombination emission packet is produced by sampling the probability that
a photon with energy larger than the ionisation threshold of H or He is emitted. 
The spectral energy distribution of the
photon packet is determined by the Milne spectrum \citep{Mihalas:1984xr}. 
After the reemission event, the cell's
counter for recombination events is put to zero and the photon packet is propagated through the box.

For further details on the algorithm and its implementation, we again refer the reader to the 
papers mentioned above.

\section{Parallelisation Strategy}
\label{sec:parStrategy} 

Monte Carlo radiation transfer methods are a powerful and easy to implement class of
algorithms that enable a determination of the radiation intensity in a simulation grid or on detectors 
(such as CCDs or photographic plates). Photons 
originating at sources are followed through the computational domain, i.e. the whole simulation box, up to 
a grid cell or the detector in a stochastic fashion \citep{Jonsson:2006jt, Juvela:2005vl, Bianchi:1996sf}. 
If only the intensity field is of interest, a straight forward parallelisation strategy is to mirror the 
computational domain and all its sources on multiple processors, so that every processor 
 holds a copy of the same data set. Then each node (a node can consist of multiple 
computational cores) propagates its own subset of the global photon sample through the
domain until the grid boundary or the detector plane is reached.
At the end, the photon counts which were determined independently on each node or core are 
gathered to the master node and are summed up to obtain the final intensity map 
\citep{J.G.-Marakis:2001ef}. This technique is also known as reduction. This strategy however only works 
if the memory requirement of the problem setup fits the memory available to each core.
What if the computational core's memory does not allow for duplication of the data? 

To solve this problem, hybrid solutions have been proposed, where the computational 
domain is decomposed {into sub-domains} and distributed to multiple 
task farms \citep{Alme:2001rp}. Each 
task farm is a collection of nodes and/or cores working on the same sub-domain. 
A task farm can either reside on just one computational node, or it can span over multiple nodes.
Each entity in the task farm propagates photons individually through 
the sub-domain until they reach the border or a detector. If photons reach the border of the 
sub-domain, they are communicated to the task farm containing the neighbouring sub-domain.
The cumulated intensity map is obtained by first aggregating the different contributions of the
computational cores in each task farm, and then by merging the
solutions of the individual task farms.
The hybrid use of distributed (multiple nodes per task farm) and shared memory concepts (domain is
shared between all cores per task farm) allows to balance the amount of 
communication that is needed between the various task farms, and the underlying computational 
complexity. 

However this method has two potential drawbacks. If, in a photon scattering 
process, the border of the sub-domain lies unfavourably in the random walk, and the photon
crosses the border multiple times in one time step, a large communication overhead is produced, 
slowing down the calculation. This eventuality arises in optically thick media. Further, in an 
optically thin medium, the mean free path of the photons can be larger than the sub-domain size. 
If the photons need to pass through a number of sub-domains during one time step, they need to
be communicated at every border crossing event. This causes a large synchronisation 
overhead. Sub-domains would need to communicate with their neighbours often per 
time step, in order to allow a synchronous propagation of photons through the domain. 
Since in {\tt CRASH2} photons are propagated instantly through the grid, each photon might
pass through many 
sub-domains, triggering multiple synchronisation events. This important issue has to be taken
into account in order to avoid inefficient parallelisation performance and scalability.

These task farm methods usually assume that the radiation intensity is determined separately from
additional physical processes, while in cosmological radiative transfer methods the interest is focused on 
the coupling of the ionising radiation transfer and the evolution of the chemical and thermal state of the gas in the IGM. 
In {\tt CRASH2}  a highly efficient algorithm is obtained
by coupling the calculation of the intensity in a cell with the evaluation of the ionisation 
network. The ionisation network is solved each
time a photon packet passes through a cell, altering its optical depth. However, if on distributed architectures 
the computational domain is copied to multiple computational nodes, one would need to 
ensure that each time the
optical depth in a cell is updated, it has to be updated on all the nodes containing a copy of the cell.
This would produce large communication overheads. Further, special care needs to be taken to
prevent two or more cores from altering identical cells at the same time, again endangering
the efficiency of the parallelisation.

One possibility to avoid the problem of updating cells across multiple nodes would be to use a 
strict shared memory approach, where a task farm consists of only one node and all the cores have
access to the same data in memory. Such an OpenMP parallelisation of the {\tt CRASH2} Monte 
Carlo method has been feasible for small problems \citep{Partl:2010wm} 
where the problem size fits one shared computational node, but would not perform well if the 
problem needs to be distributed over multiple nodes. However the unlikely case of multiple cores
accessing the same cell simultaneously still remains with such an approach.

We have decided to use the more flexible approach, by
parallelising {\tt CRASH2} for distributed memory machines using the MPI library\footnote{http://www.mpi-forum.org}. 
Using a distributed MPI approach however
requires the domain to be decomposed into sub-domains. Each sub-domain is assigned to only one
core, which means that the sub-domains become rather small when compared to the task farm
approach. Since in a typical simulation the 
photon packets will not be homogeneously distributed, the domain decomposition needs
to take this into account, otherwise load imbalances dominate over performance. This can
be addressed by adaptive load balancing, which is technically complicated to achieve, though. 
An alternative approach is to statically decompose the grid using an initial guess of the expected 
computational load. This is the method we adopted for {\tt pCRASH2}.

In order to optimise the {\tt CRASH2} code basis for larger problem sizes, the routines in
{\tt CRASH2} handling the reemission of recombination radiation had to be adapted.
Since {\tt pCRASH2} greatly extends the maximum number of photon packets that can be
efficiently processed by the code, we revert changes introduced in the recombination 
module of the serial version {\tt CRASH2} to reduce the execution time. 
To handle recombination radiation in {\tt CRASH2} effectively, photon packets produced by the
diffuse component were only emitted at fixed time steps. At these specific time steps, the whole grid 
was searched for cells that fulfilled the reemission criterium and reemission packets were emitted.
This approach allowed the sampling resolution with which the diffuse radiation
field was resolved to be controlled, depending on the choice of the recombination emission time interval. 

Searching the whole grid for reemitting cells becomes a bottle neck when larger problem sizes are
considered. In {\tt pCRASH2} we therefore reimplemented the original prescription for recombination 
radiation as described in the Maselli et al. (2003) paper. Diffuse photons are emitted whenever a cell is 
crossed by a photon packet and the recombination threshold in the cell has been reached. This results
in a more continuous emission of recombination photons and increases the resolution with which this
diffuse field is sampled. The two methods converge when the time interval between 
reemission events in {\tt CRASH2} is chosen to be very small.

\subsection{\label{sec:domDecomp} Domain decomposition strategy}

\begin{figure}
\includegraphics[width=84mm]{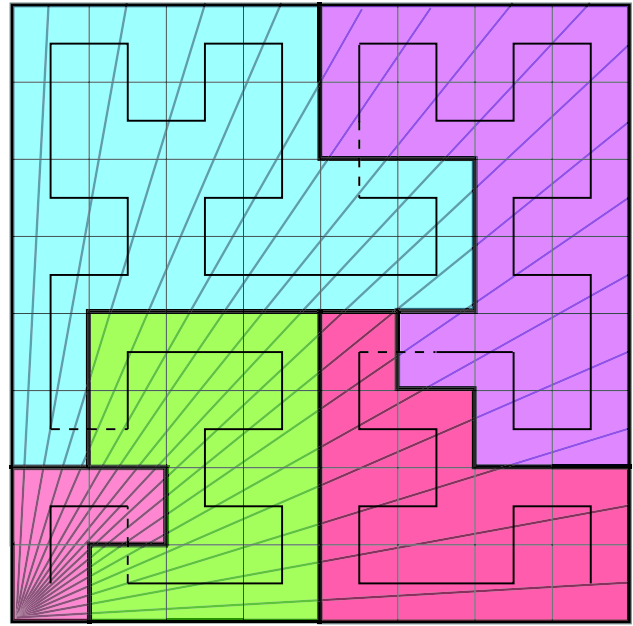}

\caption{Schematic view of the decomposition strategy with one source in the lower left corner.
The ray density decreases with $r^{-2}$ from the source. The Hilbert curve (black) is integrated 
until the threshold of the expected workload for each core is reached and the domain is cut 
(indicated by dashed lines). See text for details.}
\label{fig:hilbertDecomp}
\end{figure}

An intuitive solution to the domain decomposition problem is to divide the grid into cubes of 
equal volume. However in {\tt CRASH2} this would result in very bad load balancing and would
deteriorate the scaling performance. The imbalance arises firstly from the large
surface through which packets need to be communicated to the neighbouring domain, resulting in
large communication overhead. A decomposition strategy that minimises the surface of the 
sub-domains reduces the amount of information that needs to be passed on to neighbouring
domains. Secondly, the main contribution to the imbalance stems from the fact 
that the sub-domain 
where the source is embedded has to process more rays than sub-domains further away, 
a problem common to all ray tracing techniques. Good scaling can thus only be achieved
with a domain decomposition strategy that addresses these two issues simultaneously.
The issue of minimal communication through small sub-domain surfaces can be solved with
the right choice of how the sub-domains are constructed. The problem of optimal computational
load can only be handled with a good work load estimator which we now want to address.

In an optically thin medium the average number of rays to be processed at radius $r$ is
proportional to the ray density $\rho_{\gamma}(r) \propto r^{-2}$, as a consequence of
geometric dilution. As we do not have an a-priori knowledge of the evolution of the optical depths distribution
throughout a given run, we estimate the ray density in the optically thin approximation and
use this as a work load estimator for the domain decomposition.

For each cell at position $r_{\rm cell}$ and each source, the expected ray density $\rho_\gamma$
\begin{equation}
\rho_{\gamma} (\mathbf{r}_{\rm cell}) = \sum_{\rm sources} | \mathbf{r}_{\rm source} - \mathbf{r}_{\rm cell} |^{-2}
\label{eq:localTau}
\end{equation}
is evaluated, where $\mathbf{r}_{\rm source}$ gives the position vector of the source and
$\mathbf{r}_{\rm cell}$ the position of the cell. This yields a direct estimate of the 
computational time spent in each cell. It has to be stressed that, if for a given run large 
portions of the volume remain optically thick, the chosen estimator does not produce an
optimal load balancing. Nevertheless for these cases it is not possible to predict the evolution
of the opacity distribution before running the radiative transfer calculation, so we retain
the optically thin approximation as a compromise.

In order to achieve a domain decomposition that requires minimum communication needs (i.e. the 
surface of the sub-domain is small which reduces the amount of information that needs to be communicated
to the neighbours) and assures the sub-domains to be locally confined (i.e. that communication
between nodes does not need to be relayed far through the cluster network over multiple nodes), 
{\tt pCRASH2} implements the widely used approach of decomposing along a Peano-Hilbert 
space filling curve \citep{Teyssier:2002xi, Springel:2005xz, Knollmann:2009mq}.
The Peano-Hilbert curve is used to map the cartesian grid from the set of normal numbers
$\mathbb{N}_0^3$ to a 
one dimensional array in $\mathbb{N}_0^1$. To construct the Peano-Hilbert curve mapping 
of the domain, we use the algorithm by \citet{Chenyang:2008vl} (see Appendix 
\ref{app:Hilbert} for details on how the Peano-Hilbert curve is constructed). Then the work 
estimator is integrated along the space-filling
curve, and the curve is cut whenever the integral exceeds $\sum_i \rho_{\gamma, i} / N_{\rm{CPU}}$.
The sum gives the total work load in the grid and $N_{\rm{CPU}}$ the number of CPUs used.
This process is repeated until the curve is partitioned into consecutive segments and the grid points in 
each segment are assigned to a subdomain. Because the segmentation of the Peano-Hilbert curve
is consecutive, the sub-domains by construction are contiguously distributed on the grid and are 
mapped to nodes that are adjacent on the MPI topology. The mapping of the sub-domains onto 
the MPI topology first
maps the sub-domains to all the cores on the node, and then continues with the next adjacent
node, filling all the cores there. The whole procedure is illustrated in Fig.\ref{fig:hilbertDecomp}.

\subsection{\label{sec:paraProp} Parallel photon propagation}

One of the simplifications used in many ray-tracing radiation transfer schemes is that 
the number of photons at any position along a ray is only governed by the absorption in all
the preceding points. This approximation is generally known as the static approximation to
the radiative transfer equation \citep{Abel:1999jt}. In {\tt CRASH2} we recursively solve eq.
\ref{eq:NumPhotInCell} until the ray exits the box or all its photons are absorbed. Depending on the 
length and time scales involved in the simulation, the ray can reach distances far greater than
$c \times \Delta t$ and the propagation can be considered to be instantaneous. In such an
instant propagation approximation, each ray can pass through multiple 
sub-domains. At each sub-domain crossing, rays 
need to be communicated to neighbouring sub-domains. In one time step, there can be multiple such 
communication events, each enforcing some synchronisation between the different 
sub-domains. Such a scheme can be efficiently realised with a hybrid characteristics method
\citep{Rijkhorst:2006kx}, but has the drawback of a large communication overhead.

\begin{figure}
\includegraphics[width=84mm]{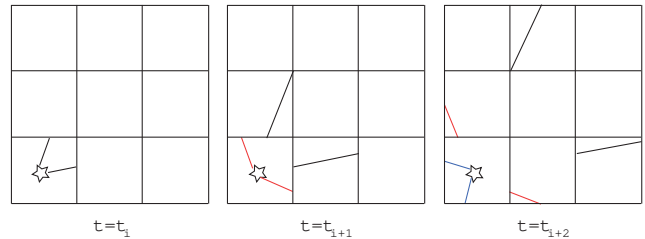}

\caption{Illustration of rays propagating through the distributed computational domain at
three time-steps $t_i, t_{i+1}$, and $t_{i+2}$. To simplify the illustration, a box domain
decomposition strategy is shown, where each square resembles one sub-domain. Further
we assume that the source only emits 
two rays per time-step. The rays are propagated to the edges of each sub-domain during 
one time step. Then they are 
passed on to the neighbouring sub-domain to be processed in the next time step.}
\label{fig:rayProp}
\end{figure}

To minimise the amount of communication phases per time step, we follow a different approach
by truncating the recursive solution of eq. \ref{eq:NumPhotInCell} at the boundaries of the
sub-domains. Instead of letting rays pass through
multiple sub-domains in one time-step, rays are only processed until they reach the boundary
of the enclosing sub-domain. At the border, they are then passed on to the 
neighbouring sub-domain. Once received, however, they will not be immediately processed by 
the neighbour. Propagation of the ray is continued in to the next time step. 
The scheme is illustrated in Fig. \ref{fig:rayProp}. In this way, each 
sub-domain needs to communicate only once per time step with all its neighbours, resulting in
a highly efficient scheme, assuming that the computational load is well distributed. 

Propagation is thus delayed and a ray needs at most $\approx 2 (N_{\rm{CPU}})^{1/3}$ 
time steps to pass through the box (assuming equipartition). The delay with which a ray is
propagated through the box depends on the size of the time-step and the number of cores
used. The larger the time-step is, the bigger the delay. The same applies for the number of cores.

If the crossing time of a ray in this scheme is well below the physical crossing time, the instant propagation 
approximation is considered retained. However rays will have differing propagation 
speeds from sub-domain to sub-domain. In the worst case scenario a ray only passes 
through one cell of a sub-domain. Therefore the minimal propagation speed is 
$v_{\rm{prop, min}} \approx 0.56 \Delta l / \Delta t$,  where $\Delta l$ is the size of one 
cell, $\Delta t$ the duration of one time step, and the factor 
0.56 is the median distance of a randomly oriented ray passing through a cell of size unity
as given in \citet{Ciardi:2001cr}. If the propagation speed of a packet 
is $v_{\rm{prop, min}} \gg c$, the propagation can be considered instantaneous, as in the
original {\tt CRASH2}. Even if this condition is not fulfilled and 
$v_{\rm{prop, min}} < c$, the resulting ionisation front and its
evolution can still be correctly modelled \citep{Gnedin:2001fc}, if the
light crossing time is smaller than the ionisation timescale, i.e. when the ionisation front propagates
at velocities much smaller than the speed of light. However the possibility exists, that near to a source, 
the ionisation timescale is shorter than the crossing time, 
resulting in the ionisation front to propagate at speeds artificially larger than light \citep{Abel:1999jt}. 
With the segmented propagation scheme it has thus to be assured that the simulation parameters are chosen
in such a way that the light crossing time is always smaller than the ionisation time scale.

The adopted parallelisation scheme can only be efficient, if the communication bandwidth per time step 
is saturated. Each time two cores need to communicate, there is a fixed overhead
needed for negotiating the communication. If the information that is transferred in one
communication event is small, the fixed overhead will dominate the communication scheme.
It is therefore important to make sure that enough information is transferred per communication
event for the overhead not to dominate the communication scheme.
The original {\tt CRASH2} scheme only allowed for the propagation of one photon packet per 
source and time-step. In order to avoid the problem described above, this restriction has 
been relaxed and each source emits multiple photon packets per time step \citep{Partl:2010wm}.
Therefore in addition to 
the total number of photon packet produced per source $N_{\rm p}$ a new simulation 
parameter governing the total number of time steps $N_{\rm{t}}$ is introduced. The number
of packets emitted by a source in one time step is thus $N_{\rm p} / N_{\rm{t}}$.

As in {\tt CRASH2}, the choice of the global time step should not exceed the smallest of the
following characteristic time scales: ionisation time scale, recombination time scale, collisional ionisation
time scale, and the cooling time scale. If this condition is not met, the integration of the ionisation network
and the thermal evolution is sub-sampled.

\subsection{\label{sec:parPRNG} Parallel pseudo-random number generators}

Since {\tt CRASH2} relies heavily on a pseudo-random numbers generator, here we have to face the challenging issue of 
generating pseudo-random numbers on multiple CPUs. Each CPU needs
to use a different stream of random numbers with equal statistical properties. However this
can be limited by the number of available optimal seeding numbers. A large collection of
parallel pseudo random number generators is available in the SPRNG library
\footnote{\url{http://sprng.cs.fsu.edu/}} \citep{Mascagni:2000cl}. From the library we are 
using the Modified Lagged Fibonacci Generator, since it provides a huge number of parallel 
streams (in the default setting of SPRNG $\approx 2^{39648}$), a large period of 
$\approx 2^{1310}$, and good quality random numbers. On top each stream returns a 
distinct sequence of numbers and not just a subset of a larger sequence.

\begin{figure}
 \includegraphics[width=84mm]{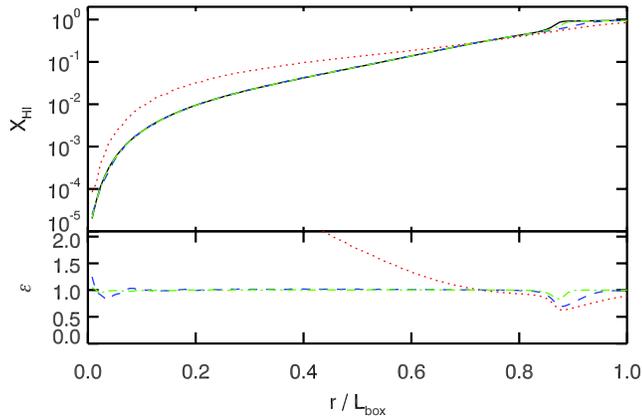}

 \caption{Test 0: Convergence study varying the number of photon packets 
 $N_{\rm{p}} = 10^6, 10^7, 10^8, 10^9$ used to simulate the evolution of an \ion{H}{ii} region. 
 Shown are the spherically averaged neutral hydrogen fraction profiles of the sphere as a function 
 of radius at the end of the simulation.
 The dotted red line gives $N_{\rm{p}} = 10^6$, the blue dashed line $N_{\rm{p}} = 10^7$, the
 green dash dotted line $N_{\rm{p}} = 10^8$, and the black solid line $N_{\rm{p}} = 10^9$.
 The test used 8 CPUs. The lower panel shows the relative deviation $\epsilon$ of the {\tt pCRASH2}
 run from the highest resolution run.}
 \label{fig:test1_conv_numPack}
\end{figure}

\begin{figure}
 \includegraphics[width=84mm]{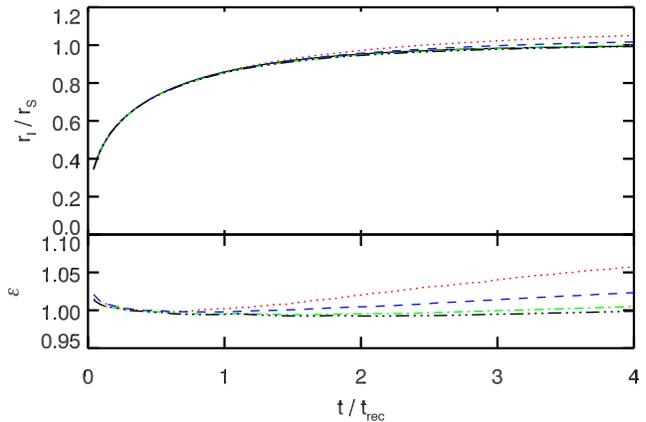}

 \caption{Test 0: Evolution of the \ion{H}{ii} region's size as a function of time normalised to the
 recombination time. The upper panel shows results for the convergence study where the number of 
 photon packets was varied. The dotted red line gives $N_{\rm{p}} = 10^6$, the blue dashed line 
 $N_{\rm{p}} = 10^7$, the green dash dotted line $N_{\rm{p}} = 10^8$, and the black dash triple-dotted
 line $N_{\rm{p}} = 10^9$. The solid black curve represents the analytic evolution $r_S(t)$.
 The test used 8 CPUs. The lower panel shows the relative deviation $\epsilon$ of the {\tt pCRASH2}
 runs from the analytic expression.}
 \label{fig:test1_conv_numPack_timeEvol}
\end{figure}

\section{Performance}
\label{sec:performance}

In this Section we present the results of an extensive comparison of the parallel
scheme with the serial one. We follow the tests described in 
\citet{Maselli:2003} and \citet{Iliev:2006fk}. Subsequently we discuss the speed performance 
of the parallel code and its scaling properties.

\subsection{Test 0: Convergence test of a pure-H isothermal sphere}
\label{sec:test0}

First we study the convergence behaviour of a Str\"omgren sphere as a function of the 
number of photon packets $N_{\rm{p}}$ and the number of time steps $N_{\rm{t}}$. The setup
of this test is equivalent to Test 1 in the code comparison project \citep{Iliev:2006fk}.
For this test, we distribute the computational domain over 8 CPUs.

The time evolution of an \ion{H}{ii} region produced by a source having a $10^5 \textrm{ K}$ black-body
spectrum with constant intensity expanding in a homogeneous medium is followed. The
hydrogen density of the homogeneous medium is fixed at 
$n_{\rm{H}} = 10^{-3} \textrm{ cm}^{-3}$, with no helium included. The temperature is fixed at
$T=10^4 \textrm{ K}$ and kept constant throughout the calculation. The initial ionisation fraction
is initialised with $x_{\ion{H}{ii}} = 1.2\times 10^{-3}$. The grid has a linear size
of $L_{\rm{box}}=6.6 \textrm{ kpc } h^{-1}$ and is composed of $N_{c}^3=128^3$ cells. 
The ionising source produces $\dot N_{\gamma} = 5\times 10^{48} \textrm{ photons s}^{-1}$. 
Recombination radiation is followed and photons are emitted whenever 10\% of
the electrons in a cell have recombined. The simulation time is set at 
$t_{s} = 5 \times 10^8 \textrm{ yr}$ which is approximately four times the recombination 
timescale.

In Fig. \ref{fig:test1_conv_numPack} we present the resulting spherically averaged profile of the 
neutral hydrogen fraction where we use $N_{\rm{p}} = 10^6, 10^7, 10^8, 10^9$ photon
packets. The number of time steps $N_{\rm{t}}$ is equal to the number of packets 
in order to obtain the identical scheme used in the serial version of {\tt CRASH2}. 
$N_{\rm{p}} = 10^6$ packets are
unable to reproduce the correct neutral hydrogen fraction profile. Using $N_{\rm{p}} = 10^7$ 
packets instead already yields satisfactory results, however we consider the solution to have
converged with at least $N_{\rm{p}} = 10^8$ packets. For this case the mean relative 
deviation $\left\langle \epsilon (r) \right\rangle = \left\langle x_{\ion{H}{i}}(r) / x_{\ion{H}{i}, \mathrm{ref}} (r) \right\rangle$ of the neutral fraction 
to the $N_{\rm{p}} = 10^9$ run is $0.7 \%$. The largest relative deviation is found
in a small region in the ionisation front itself. Its width is sensitive to the sampling resolution, 
where the solution for the $N_{\rm{p}} = 10^8$ run differs by $18\%$ from the
$N_{\rm{p}} = 10^9$ run. In Fig. \ref{fig:test1_conv_numPack_timeEvol} the time evolution of the
\ion{H}{ii} region's size is given. The radius is calculated using the volume of the \ion{H}{ii} region, which
is determined with $V = \sum_\mathrm{cells} x_{\ion{H}{ii}, i} (\Delta l)^3$, where $\Delta l$ denotes
the physical width of one cell. {\tt pCRASH2} is able to resolve the time evolution of the \ion{H}{ii} region
up to $2.5\%$ accuracy for $N_{\rm{p}} = 10^7$ when compared with the analytic expression $r_{\mathrm{I}}(t)
= r_{\mathrm{S}} \left( 1 - \exp(-t/t_{\mathrm{rec}}) \right)$, where $t_{\mathrm{rec}}$ is the recombination
time. Deviations from the analytic solution at $4 t_{\mathrm{rec}}$ decrease to 0.5\% for 
$N_{\rm{p}} = 10^8$ and $0.2\%$ for $N_{\rm{p}} = 10^9$.
Because of these findings, we use $N_{\rm{p}} = 10^9$ photon packets
whenever the serial version allows for such a high sampling resolution, otherwise 
$N_{\rm{p}} = 10^8$ packets will be used. We further note that the impact of the sampling
resolution on the structure of the ionisation front is very important when the formation and destruction
of molecular hydrogen is considered. It can significantly affect the stability of the ionisation front
in cosmological simulations, in particular by introducing inhomogeneities in the star formation \citep{Ricotti:2002fr,
Susa:2006mz, Ahn:2007gf}.

\begin{figure}
 \includegraphics[width=84mm]{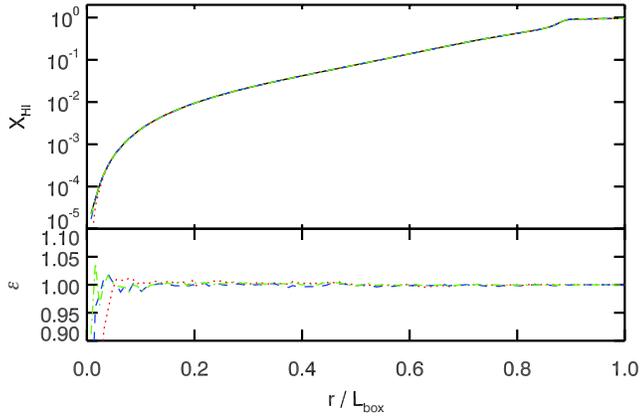}

 \caption{Test 0: Convergence study varying the number of time steps used to simulate the  
 evolution of an \ion{H}{ii} region. 
 Shown are the spherically averaged neutral hydrogen fraction profile of the sphere as a function 
 of radius at the end of the simulation.
 The dotted red line gives the results using $N_{\rm{t}} = 10^5$ time steps, the blue dashed one
 $N_{\rm{t}} = 10^6$ time steps, the green dash dotted one $N_{\rm{t}} = 10^7$, and the black
 solid line $N_{\rm{t}} = 10^8$ time steps. A constant number of $N_{\rm{p}} = 10^8$ has been 
 used.  The test was run on 8 CPUs. The lower panel shows the relative deviation $\epsilon$
 of the {\tt pCRASH2} run from the highest time resolution run.
 }
 \label{fig:test1_conv_numTime}
\end{figure}

\begin{figure}
 \includegraphics[width=84mm]{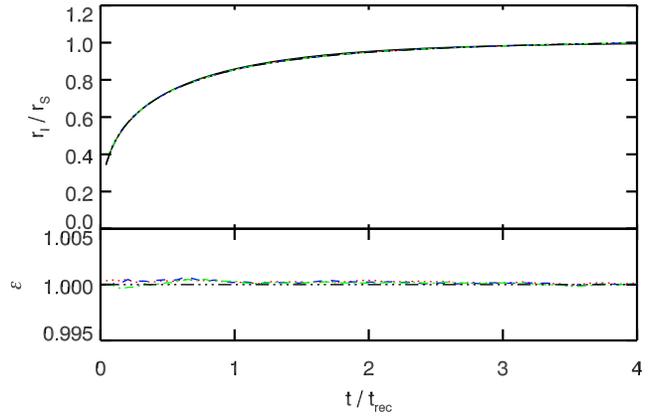}

 \caption{Test 0: Evolution of the \ion{H}{ii} region's size as a function of time normalised to the
 recombination time. The upper panel shows results for the convergence study where the number of 
 time steps used in the simulation was varied. The number of packets used was fixed at 
 $N_{\rm{p}}=10^8$. 
 The dotted red line gives the results using $N_{\rm{t}} = 10^5$ time steps, the blue dashed one
 $N_{\rm{t}} = 10^6$ time steps, the green dash dotted one $N_{\rm{t}} = 10^7$, and the black
 dashed triple-dotted line $N_{\rm{t}} = 10^8$ time steps.
 The solid black curve represents the analytic evolution $r_S(t)$.
 The lower panel shows the relative deviation $\epsilon$ of the {\tt pCRASH2} runs from the 
 solution using $N_{\rm{t}} = 10^8$ time steps.}
 \label{fig:test1_conv_numTime_timeEvol}
\end{figure}

For the test where the number of time steps is varied (and thus the number of photons emitted
per time step), we use $N_{\rm{p}} = 10^8$ packets. The lower sampling resolution is
used to reduce the accuracy with which the ionisation network is solved. 
Any influence of the time step length should be more easily seen if the
accuracy in the solution of the ionisation network is lower.   
We evolve the \ion{H}{ii} region using
$N_{\rm{t}} = 10^5, 10^6, 10^7, 10^8$ time steps. The resulting neutral hydrogen fraction
profiles are given in Fig. \ref{fig:test1_conv_numTime}. Overall, varying the number of time steps
does not alter the resulting profile significantly. Only in the direct vicinity of the source, the solution
is sensitive to the number of time steps used. In Fig. \ref{fig:test1_conv_numTime_timeEvol}
the time evolution of the \ion{H}{ii} region's radius is shown for this test. The fluctuations between
the results for various numbers of time steps are small, when compared to the $N_{\rm{t}} = 10^8$. 
This is especially true for the equilibrium solution. At early times when the \ion{H}{ii} region grows rapidly,
the largest deviations can be seen. However for none of the cases the difference exceeds $0.05\%$.
Since the choice of $N_{\rm{t}}$ 
directly determines the propagation speed of the photon packets, we confirm that the choice
of propagation speed is not affecting the resulting \ion{H}{ii} region and that the new ray propagation
scheme is a valid approach.

\begin{figure}
 \includegraphics[width=84mm]{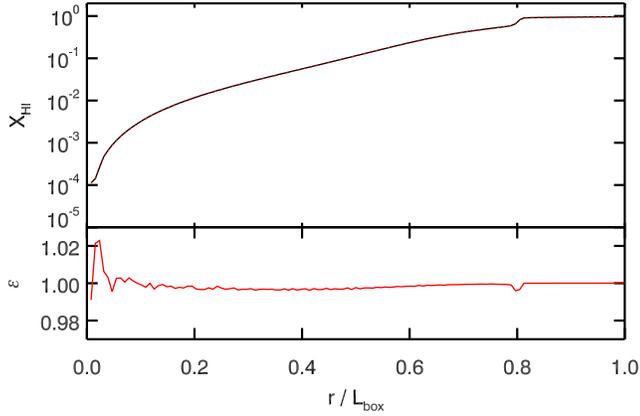}

 \caption{Test 1: Comparing the solvers of {\tt pCRASH2} with {\tt CRASH2}.  Shown are the spherically 
 averaged neutral hydrogen fraction profile of the sphere as a function of radius at the end of 
 the simulation. The red dotted line gives the {\tt pCRASH2} results and the black solid line the
 solution obtained with {\tt CRASH2}. {\tt pCRASH2} was run on 8 CPUs. 
 The lower panel shows the
 relative deviation $\epsilon$ of the {\tt pCRASH2} run from the serial {\tt CRASH2} run.}
 \label{fig:testLuca_noRecomb}
\end{figure}

\subsection{Comparing pCRASH2 with CRASH2}
\label{sec:tests}

\subsubsection{Test 1: Pure-H isothermal sphere}
\label{sec:test1}

Now we compare the results obtained from {\tt pCRASH2} with those from the serial {\tt CRASH2} code. 
Since {\tt pCRASH2} evolves recombination radiation smoother (remember {\tt CRASH2} emits 
recombination photons at specific time steps while {\tt pCRASH2} emits diffuse radiation 
continuously), we expect differences between the two codes in regions where 
recombination radiation dominates. In order to show that {\tt pCRASH2}'s solver performs 
identically to the one in {\tt CRASH2}, we use the same setup
as in Test 0, but we do not allow for recombination radiation to be emitted. For both codes, 
we set $N_{\rm{p}} = 10^9$. The {\tt pCRASH2} solution is obtained on 8 CPUs, using 
$N_{\rm{t}} = 10^7$ time steps.

The resulting neutral hydrogen fraction profile is shown in Fig. \ref{fig:testLuca_noRecomb}.
The two curves of the {\tt CRASH2} and the {\tt pCRASH2} runs are barely distinguishable. 
The fluctuations in the relative deviations from the {\tt CRASH2} 
results near to the source are due to fluctuations in the least significant decimal.
Only at the ionisation front, a small deviation of around 0.5\% is seen. 
In the convergence study we have seen that the radial profile is sensitive to the sampling 
resolution at the ionisation front (IF) where the partially ionised medium abruptly turns into 
a completely neutral one. It is as well sensitive to the variance introduced by
using different random number generator seeds. Tests using different seeds have shown, that
fluctuations of up to $0.4\%$ between the various results can be seen. Therefore the small deviation in the
ionisation front stems to some extent from the fact that {\tt pCRASH2} uses a different set of random numbers.
The parallel solver thus gives results which are in agreement with the serial version. 
Including recombination 
radiation however will chance this picture, which we address with the next test. 

\begin{figure*}
 \includegraphics[width=84mm]{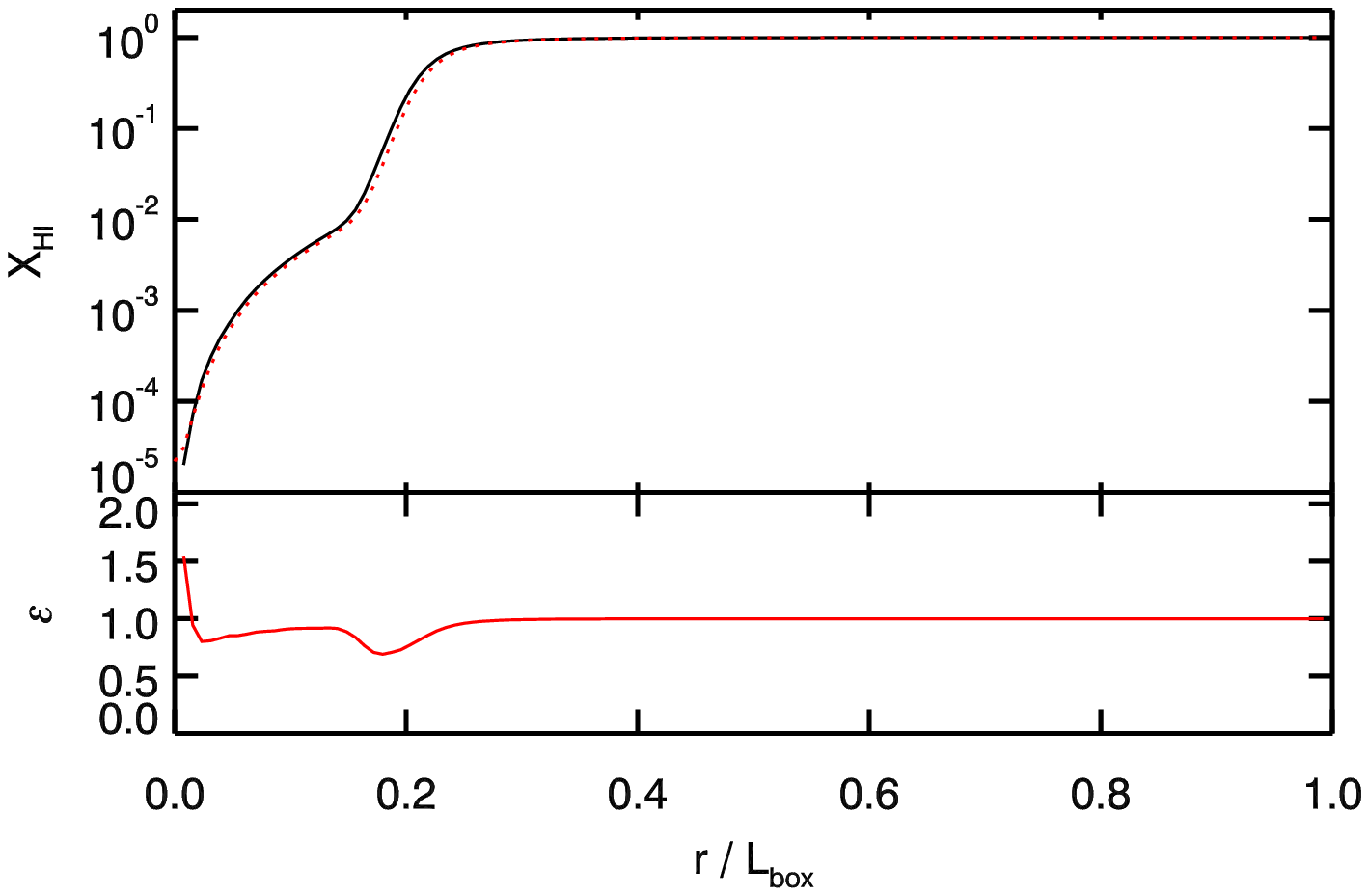}
 \hspace{5mm}
 \includegraphics[width=84mm]{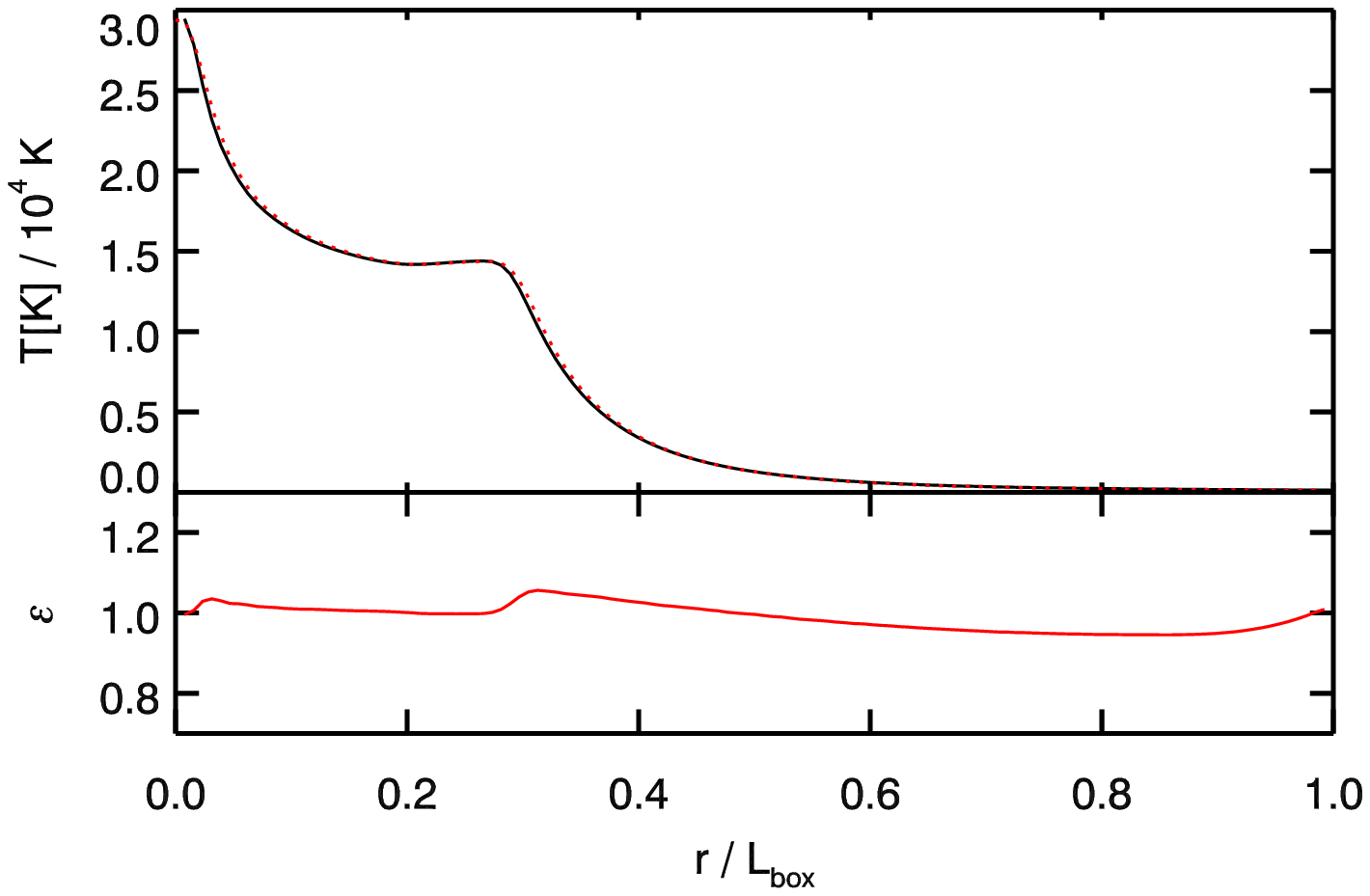} \\

 \includegraphics[width=84mm]{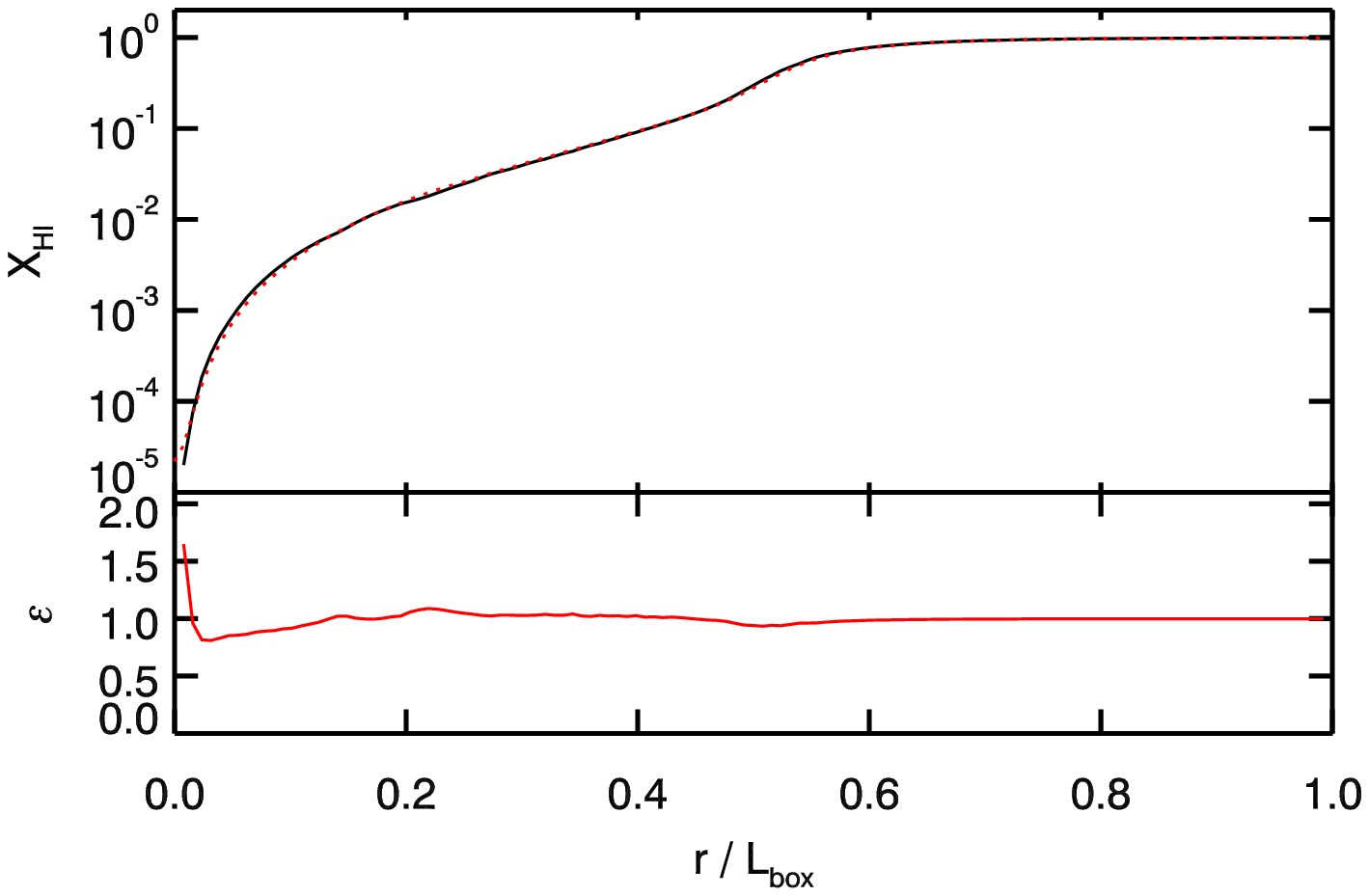}
 \hspace{5mm}
 \includegraphics[width=84mm]{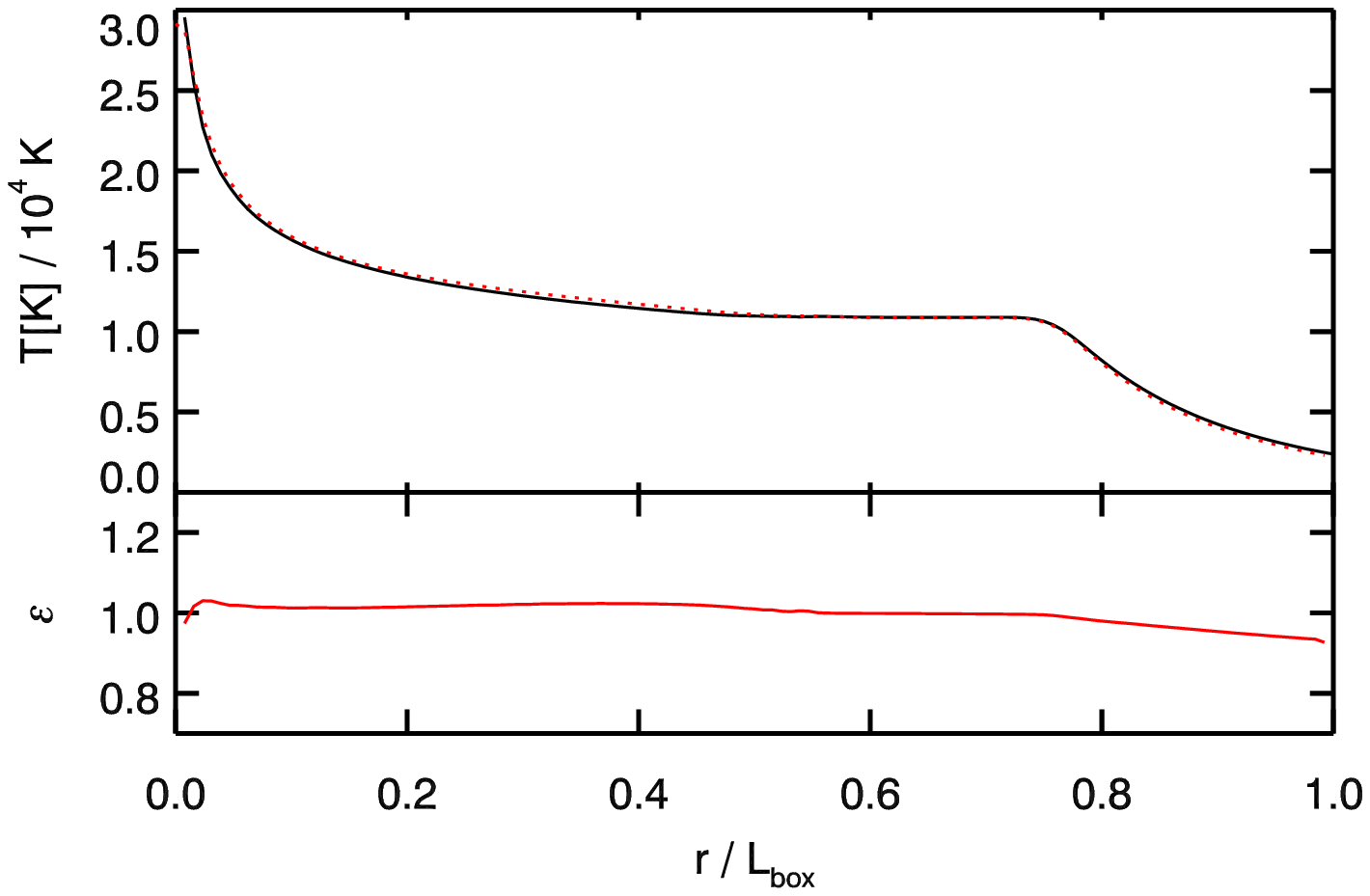} \\

 \includegraphics[width=84mm]{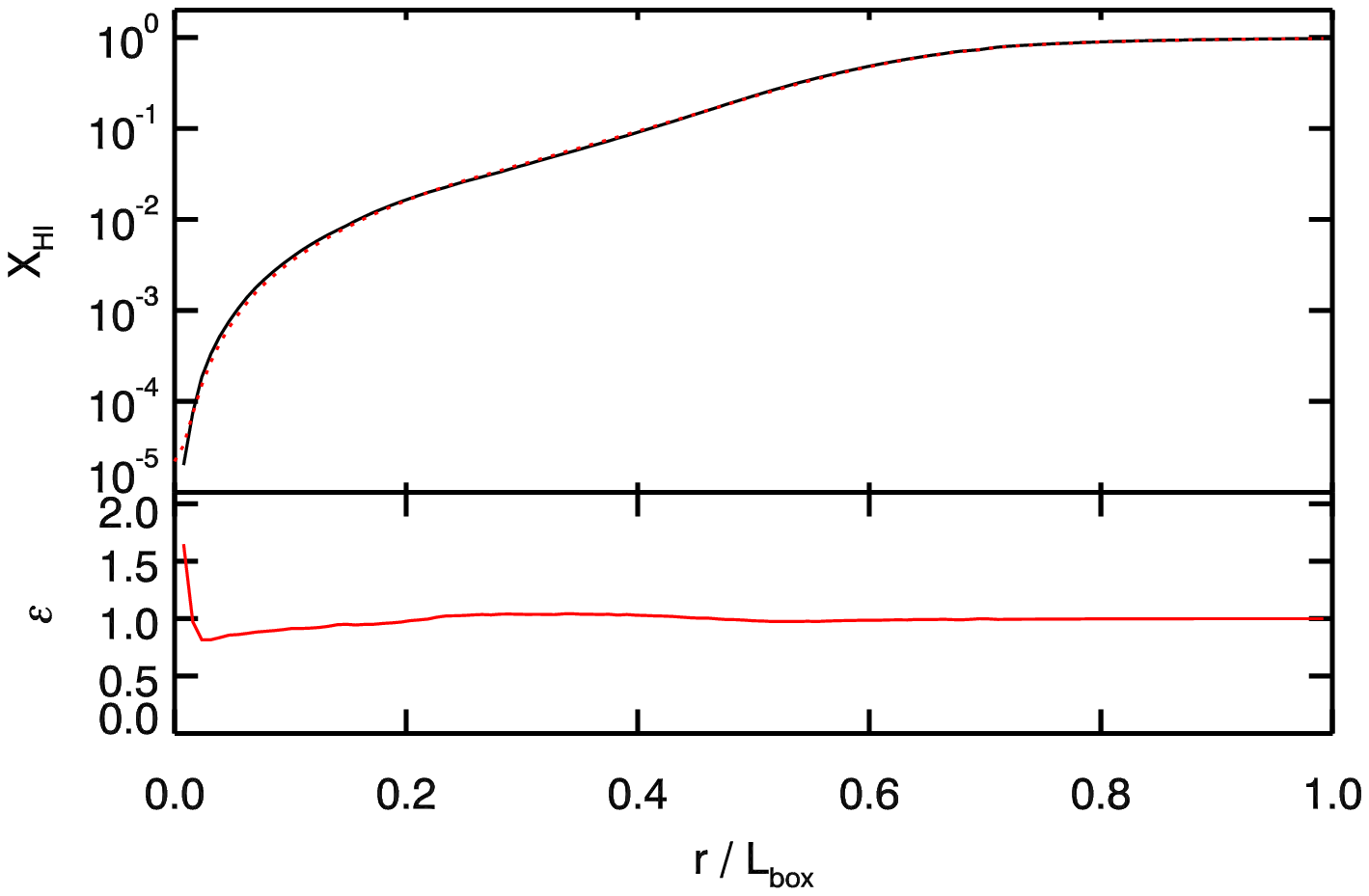}
 \hspace{5mm}
 \includegraphics[width=84mm]{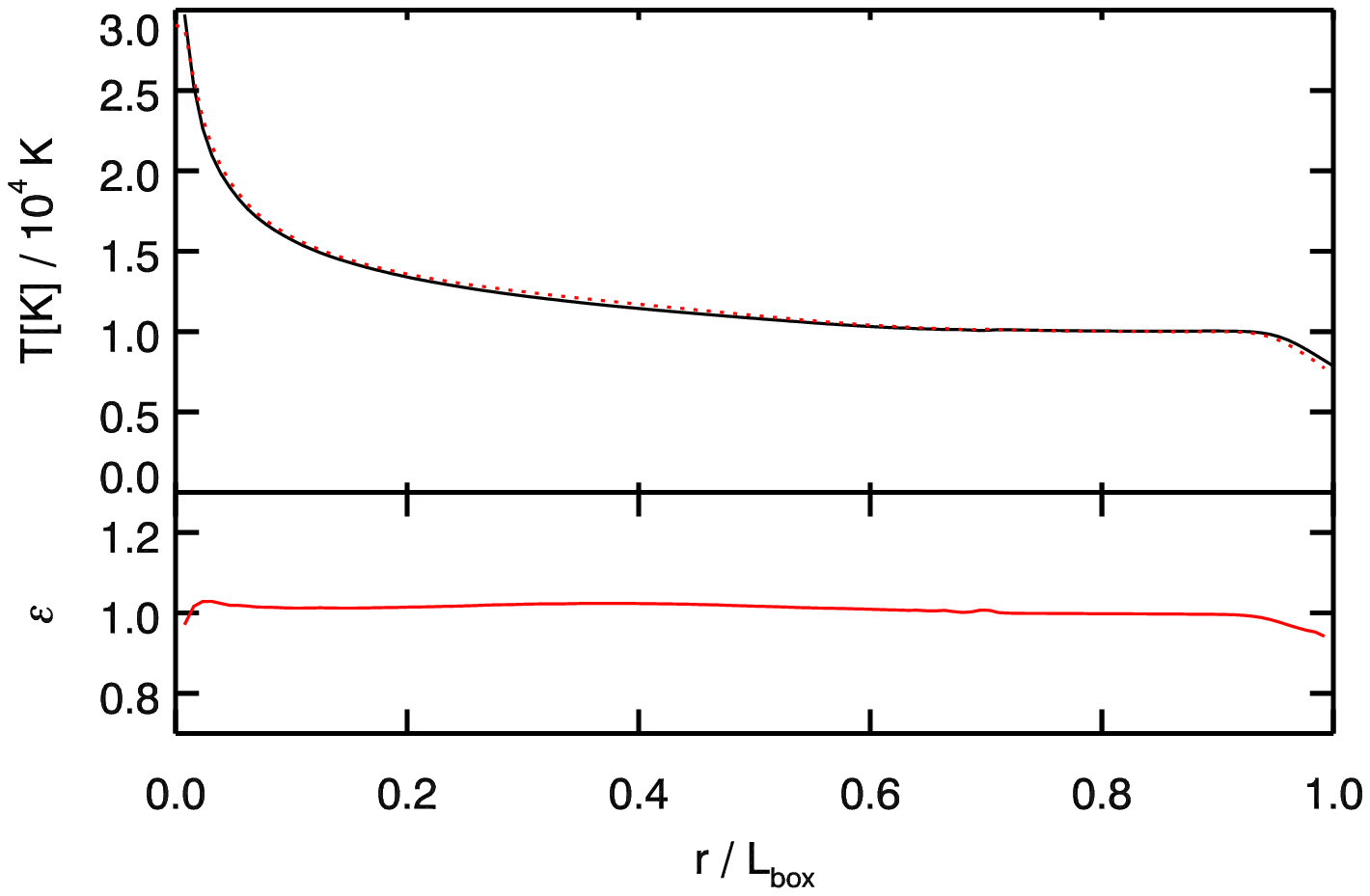} \\

 \caption{Test 2: Comparing the results of {\tt pCRASH2} (red dotted line) with {\tt CRASH2} 
 (black line). Shown are the spherically averaged neutral hydrogen fraction profiles of the sphere 
 as a function of radius (left column) and the temperature profiles (right column) at 
 $t=1\times10^7, 2\times10^8, 5\times10^8 \textrm{ yr}$ (from top to bottom). 
 The {\tt pCRASH2} results are obtained on 8 CPUs. 
 The small bottom panels show the relative deviation from the {\tt CRASH2} reference solution.}
 \label{fig:test2CompProfiles}
\end{figure*}

\subsubsection{Test 2: Realistic \ion{H}{ii} region expansion.}
\label{sec:test2}

The test we are now focusing on is identical to Test 2 in the radiative transfer codes 
comparison paper \citep{Iliev:2006fk}, except that we use a larger box size to better accommodate
the region preceding the ionisation front where preheating due to higher energy photons is important.
We use the same setup as in Test 0, but now we follow the temperature evolution, and 
self-consistently account for the reemission. Again the gas is initially fully neutral. Its initial
temperature however is now set to $T = 100 \textrm{ K}$. The source emits 
$N_{\rm{p}} = 4 \times 10^9$ photon packets.  Again we compare the solution
obtained with {\tt CRASH2} to the one obtained with {\tt pCRASH2}, which for the present test is run with 
$N_{\rm{t}} = 10^7$ time steps on 8 CPUs.

In Fig. \ref{fig:test2CompProfiles} we show the resulting neutral hydrogen fraction profiles
and temperature profiles at three different time steps 
$t=1\times10^7, 2\times10^8, 5\times10^8 \textrm{ yrs}$. Plotted are the solutions obtained
with {\tt CRASH2} and with {\tt pCRASH2} (solid black and red dotted lines respectively). At $t=1\times10^7 \textrm{ yrs}$, the
results of {\tt CRASH2} and {\tt pCRASH2}  slightly differ at the ionisation front. The structure
of the ionisation front appears slightly sharper in the {\tt pCRASH2} run, than in
the {\tt CRASH2} run. In the front, recombination radiation is an important process
\citep{Ritzerveld:2005}. Since {\tt pCRASH2} continuously emits recombination photons 
whenever a cell reaches the recombination threshold and not 
only at fixed time steps, the diffuse field is better sampled and the ionisation front becomes
sharper. For this test, {\tt CRASH2}
sampled the diffuse field $10^3$ times through the whole simulation. In {\tt pCRASH2} the
number of cell crossings determines the sampling resolution of the recombination radiation.
Therefore recombination radiation originating in the cells of the ionisation front is evaluated
at least $5 \times 10^4$ times for this specific run.
A higher sampling rate in the recombination radiation reproduces
its spectral energy distribution which is strongly peaked towards the \ion{H}{i} 
photo-ionisation threshold (see the Milne spectrum) more accurately, resulting in a sharper 
ionisation front. Better sampling of the Milne spectrum reproduces its shape more accurately
and reduces the possibility that too much energy is deposited in the high energy tail of the
Milne spectrum due to the poor sampling resolution of the spectrum.

\begin{figure}
 \includegraphics[width=84mm]{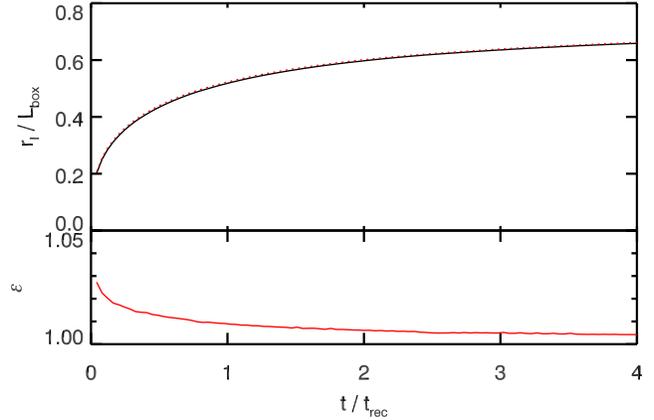}

 \caption{Test 2: Comparing the time evolution of the \ion{H}{ii} region's radius from the {\tt pCRASH2} run
 (red dotted line) with {\tt CRASH2} (black line). The {\tt pCRASH2} results are obtained on 8 CPUs. 
 The small bottom panel show the relative deviation from the {\tt CRASH2} reference solution.}
 \label{fig:test2CompTimeEvol}
\end{figure}

At later times the two codes give similar results for the hydrogen ionisation fraction, 
with almost negligible differences close to the source and across the ionisation front. 
 This is also seen in the temporal evolution of the \ion{H}{ii} region's radius given in Fig. 
\ref{fig:test2CompTimeEvol}, which is determined
as described in Test 0. In the beginning the differences between the codes are at most $3\%$, 
decreasing steadily up to one recombination time. This clearly shows the effect of the smooth
emission of recombination radiation in {\tt pCRASH2}. After one recombination time, the differences
between the two codes are small and convergence is achieved. A similar picture
emerges for the temperature profile. As for the ionisation front, the preceding temperature 
front is as well slightly sharper with {\tt pCRASH2}.

\subsubsection{Test 3: Realistic \ion{H}{ii} expansion in a H+He medium}
\label{sec:testCC}

In this test, we study the expansion of an \ion{H}{ii} region in a medium composed of hydrogen
and helium. This corresponds to the CLOUDY94 test in \citet{Maselli:2009gd}. We consider
the \ion{H}{ii} region produced by a $T=6\times10^4 \textrm{ K}$ blackbody radiator with a
luminosity of $L=10^{38} \textrm{ erg s}^{-1}$. The point source ionises a uniform medium
with a density of $n=1 \textrm{ cm}^{-3}$ in a gas composed of 90\% hydrogen
number density and the rest helium. The temperature is initially set at $T=10^2\textrm{ K}$ 
and its evolution is solved for. The dimension of the box is $L_\mathrm{box} = 128 \textrm{ pc}$. 
The test is run for $t_{s}=6 \times 10^5 \textrm{ yrs}$ after which
ionisation equilibrium is reached with $N_{\rm{p}} = 2 \times 10^8$. {\tt pCRASH2} was run on 
8 CPUs with $N_{\rm{t}} = 10^8$.

\begin{figure*}
 \includegraphics[width=84mm]{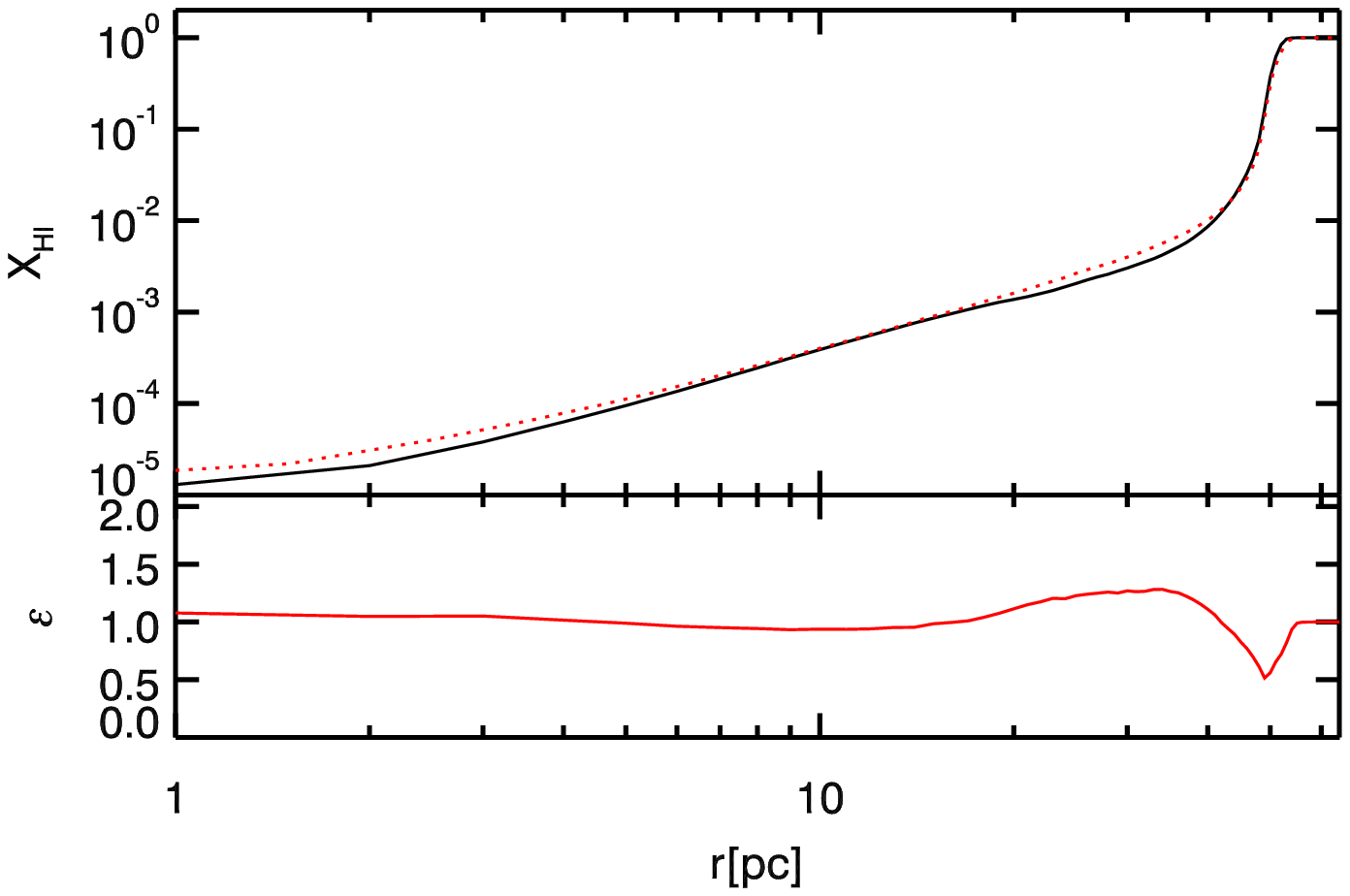} 
 \hspace{5mm}
 \includegraphics[width=84mm]{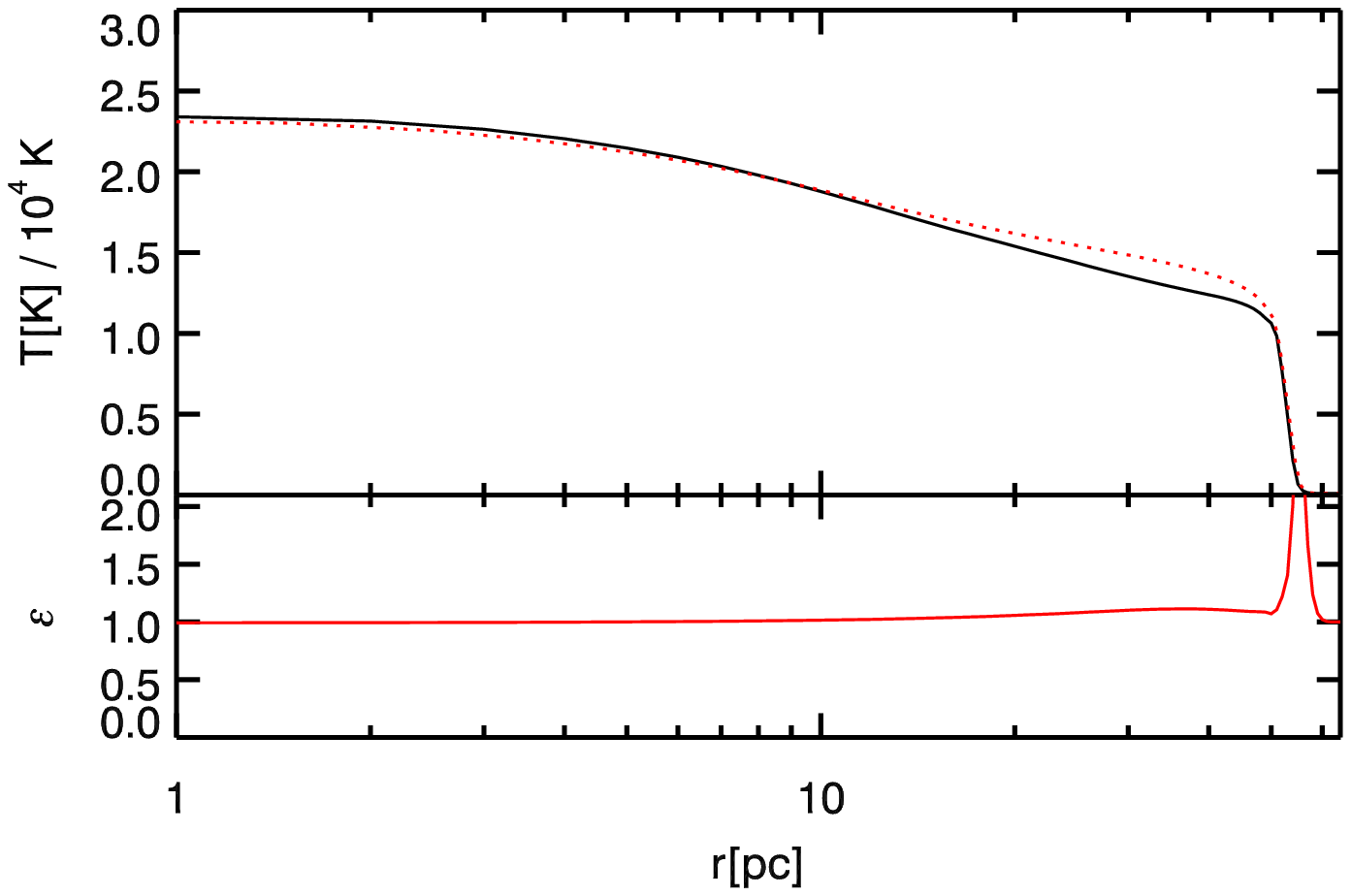} \\

 \includegraphics[width=84mm]{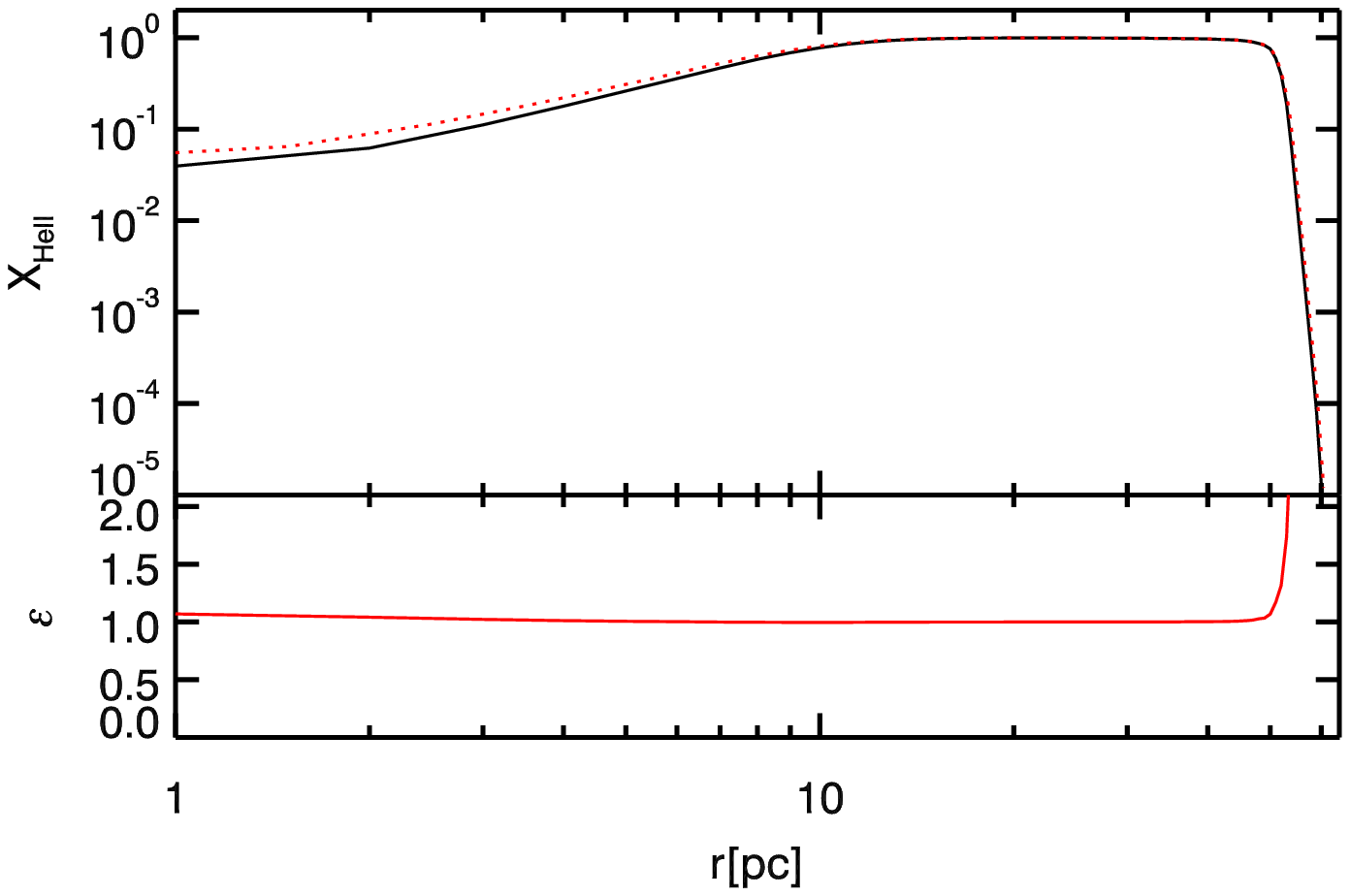}
 \hspace{5mm}
 \includegraphics[width=84mm]{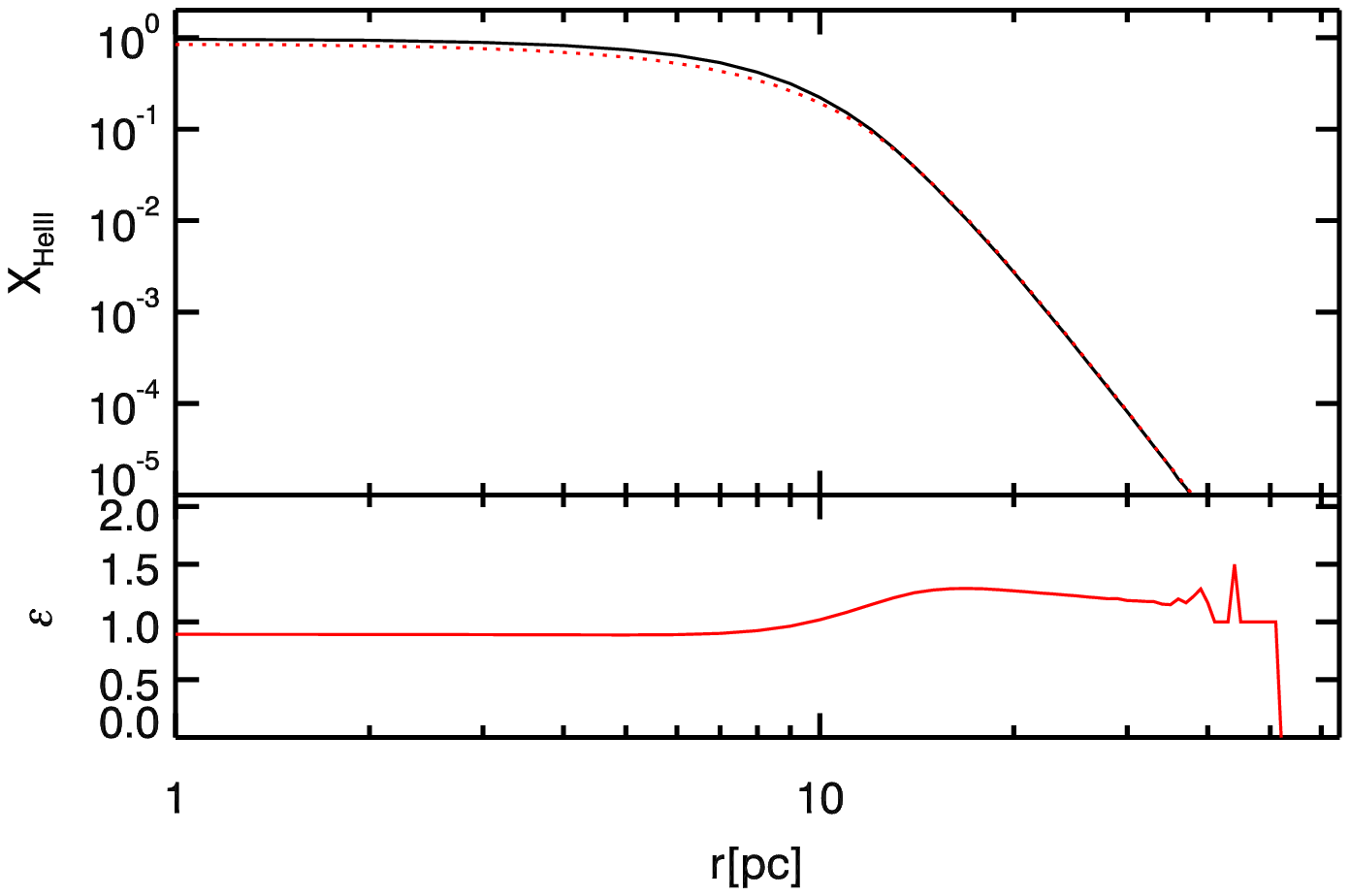} \\

 \caption{Test 3: Realistic \ion{H}{ii} region extending in a hydrogen + helium medium. 
 Compared are equilibrium results obtained with {\tt CRASH2} (black solid line) to ones obtained with 
 {\tt pCRASH2} (red dotted line). The upper left 
 panel gives the profile of neutral hydrogen fractions, lower left panel gives singly ionised helium
 fractions, and the lower right panel gives the double ionised helium fractions. The upper right
 panel gives the spherically averaged temperature profile as a function of radius.The small bottom
 panels show the relative deviation of the {\tt pCRASH2} run from the {\tt CRASH2} reference solution.}
 \label{fig:testCC_profiles}
\end{figure*}

In Fig. \ref{fig:testCC_profiles} we compare the resulting temperature, hydrogen and helium 
density profiles from {\tt pCRASH2} with the ones obtained with {\tt CRASH2}. Overall the resulting
{\tt pCRASH2} profiles are in good agreement with {\tt CRASH2}. Again only slight deviations 
from the {\tt CRASH2} solution are present in the outer parts of the \ion{H}{ii} region due to higher
sampling resolution of the ionising radiation re-emitted in recombinations. As in the previous test, the {\tt pCRASH2}
solution shows somewhat sharper fronts in hydrogen and temperature. The hydrogen fronts
are shifted slightly further away from the source in {\tt pCRASH2}, giving rise to the large 
relative error. Otherwise {\tt pCRASH2}'s performance is equivalent to that of {\tt CRASH2}. 

\subsubsection{Test 4: Multiple sources in a cosmological density field}
\label{sec:test4}

The setup of this test is identical to Test 4 in the radiative transfer codes 
comparison paper \citep{Iliev:2006fk}. Here the formation of \ion{H}{ii} regions from multiple
sources is followed in a static cosmological density field at redshift $z=9$, including
photo-heating. The initial temperature is set to $T=100 \textrm{ K}$. The positions
of the 16 most massive halos are chosen to host $10^5 \textrm{ K}$ black-body radiating 
sources. Their luminosity is set to be proportional to the corresponding halo mass and all sources are assumed to
switch on at the same time. No periodic boundary conditions are used. 
The ionisation fronts are evolved for $t_{s}=4 \times 10^7 \textrm{ yr}$. Each source produces
$N_{\rm{p}} = 1 \times 10^7$ photon packets. {\tt pCRASH2} is run with 
$N_{\rm{t}} = 10^6$ time steps on 16 CPUs.

\begin{figure*}
 \includegraphics[width=84mm]{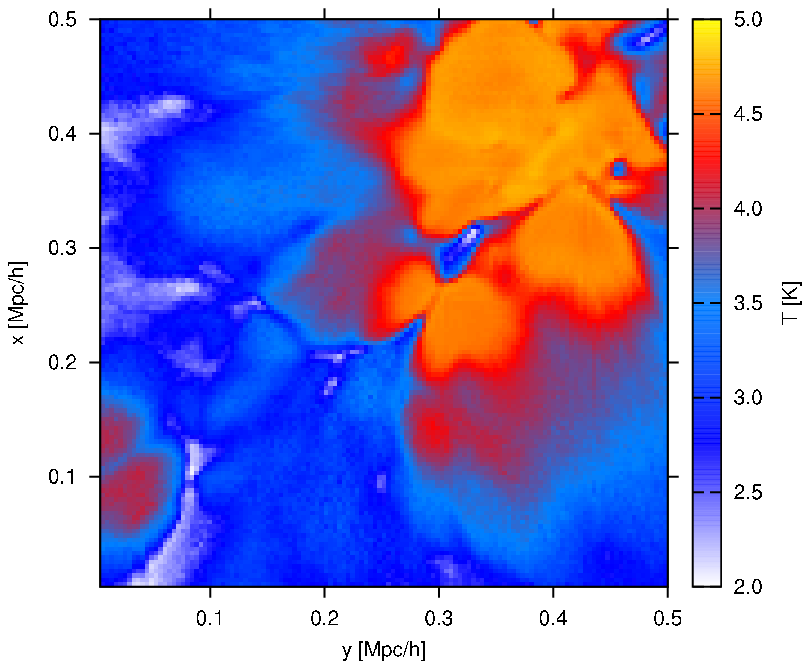}
 \hspace{5mm}
 \includegraphics[width=84mm]{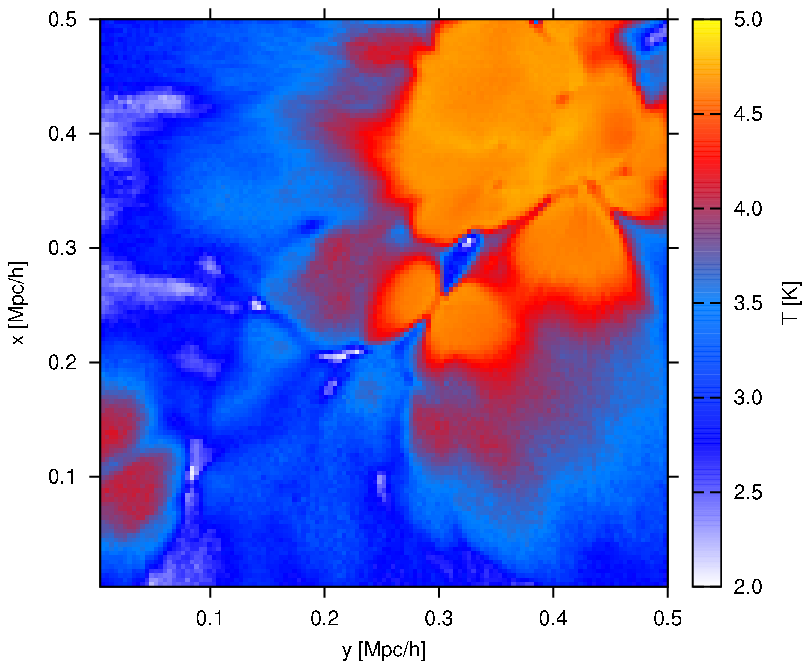} \\

 \includegraphics[width=84mm]{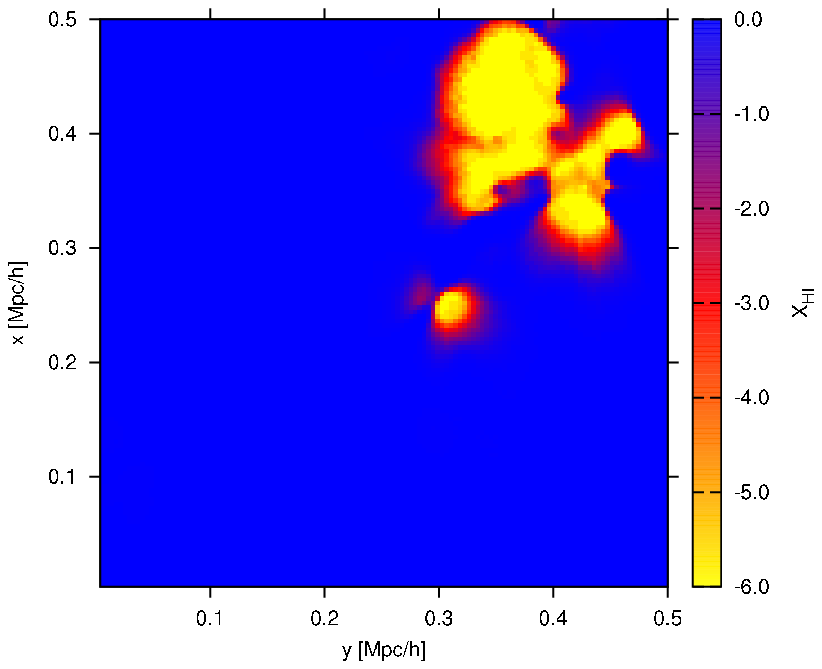}
 \hspace{5mm}
 \includegraphics[width=84mm]{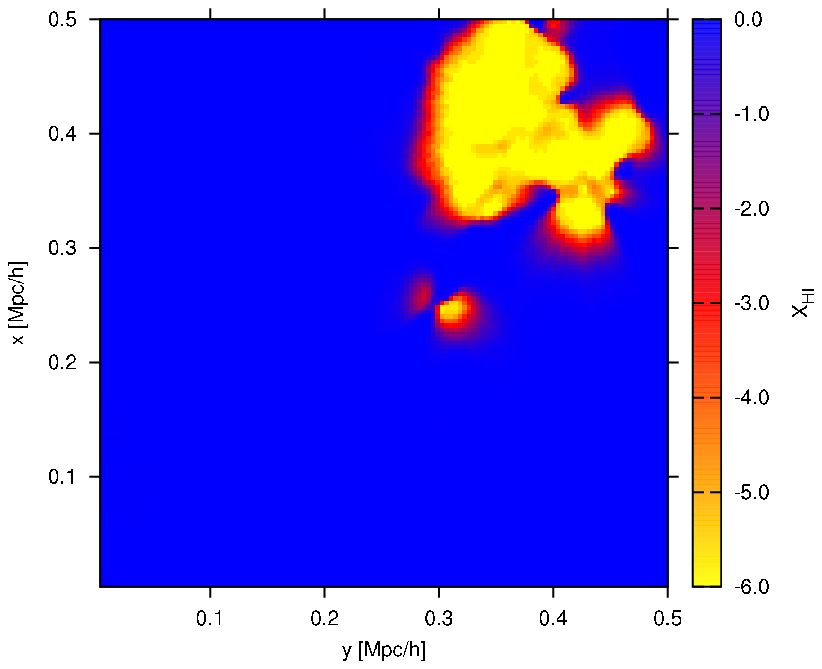} \\

 \includegraphics[width=84mm]{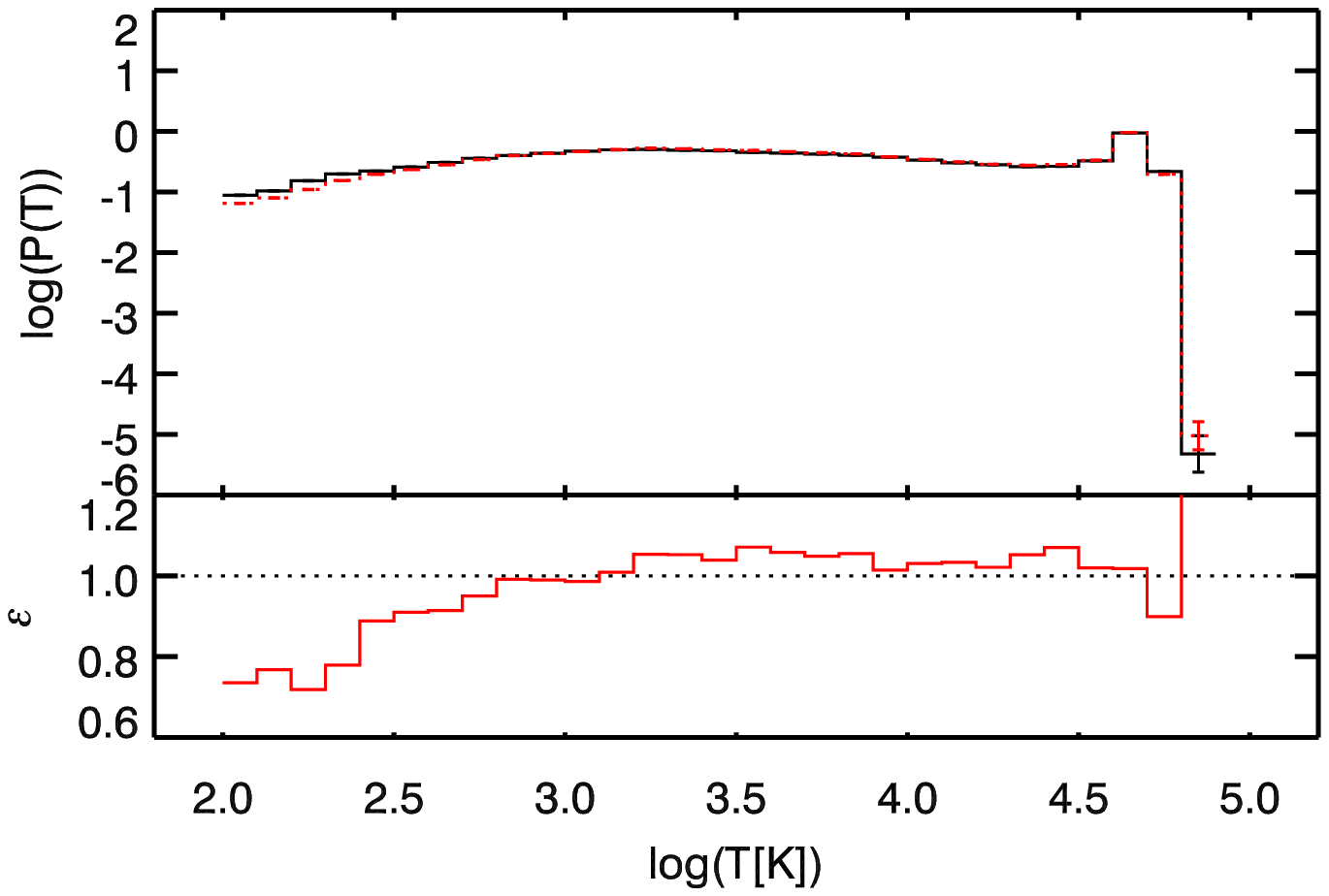}
 \hspace{5mm}
 \includegraphics[width=84mm]{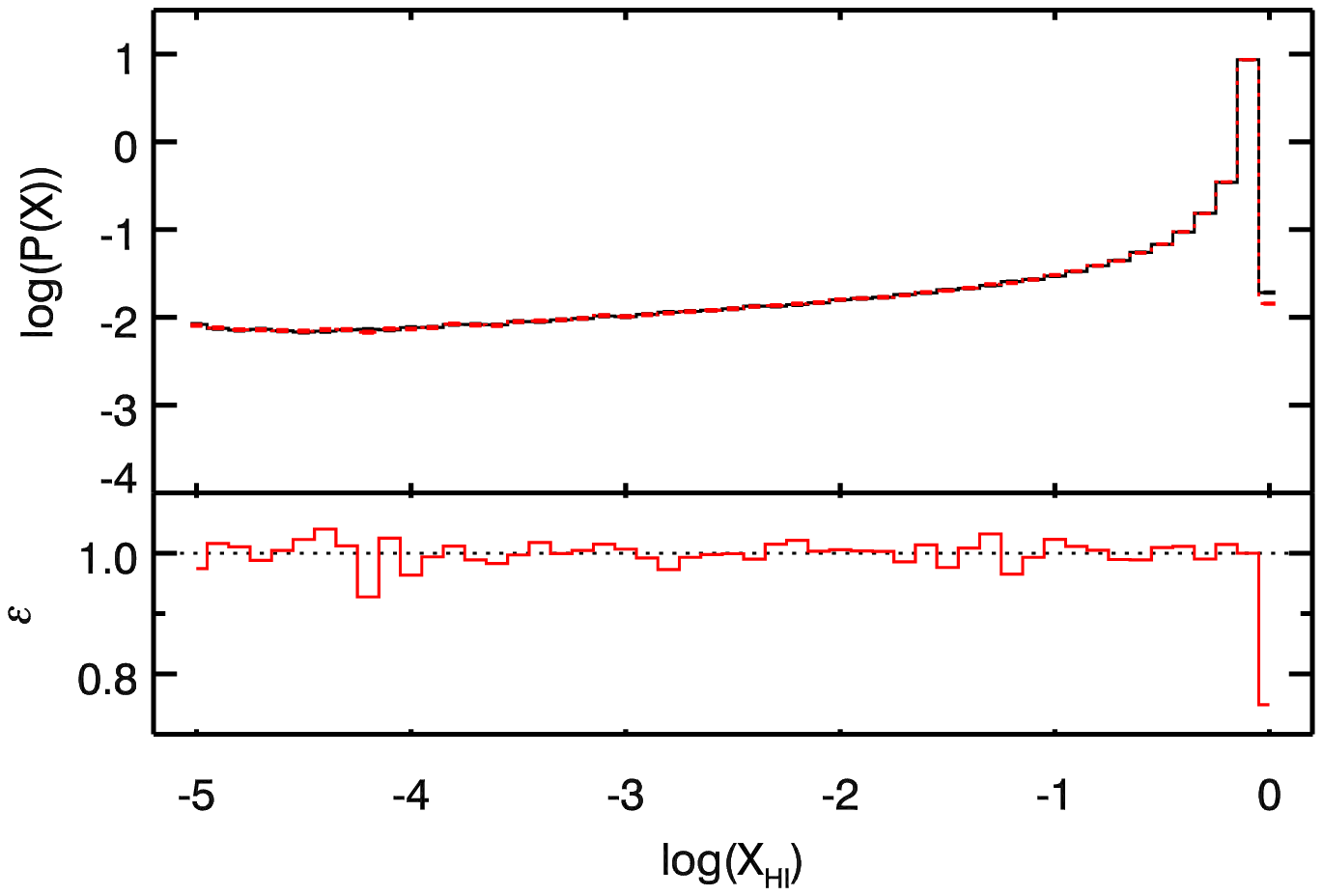} \\

 \caption{Test 4: Cut through the temperature fields (upper panels) and neutral hydrogen 
 fractions (middle panels) at time-step $t=0.05 \textrm{ Myr}$. Left panels show the solution 
 obtained with {\tt CRASH2}, right panels show the {\tt pCRASH2} solution. The lower panels give 
 volume weighted temperature and neutral fraction probability distribution functions obtained with {\tt CRASH2} 
 (black) and {\tt pCRASH2} (red dotted), where the error bars give Poissonian errors. 
 The small bottom panels show the relative deviation of the
 {\tt pCRASH2} run from the 
 {\tt CRASH2} reference solution.}
 \label{fig:test4_time-step1}
\end{figure*}

\begin{figure*}
 \includegraphics[width=84mm]{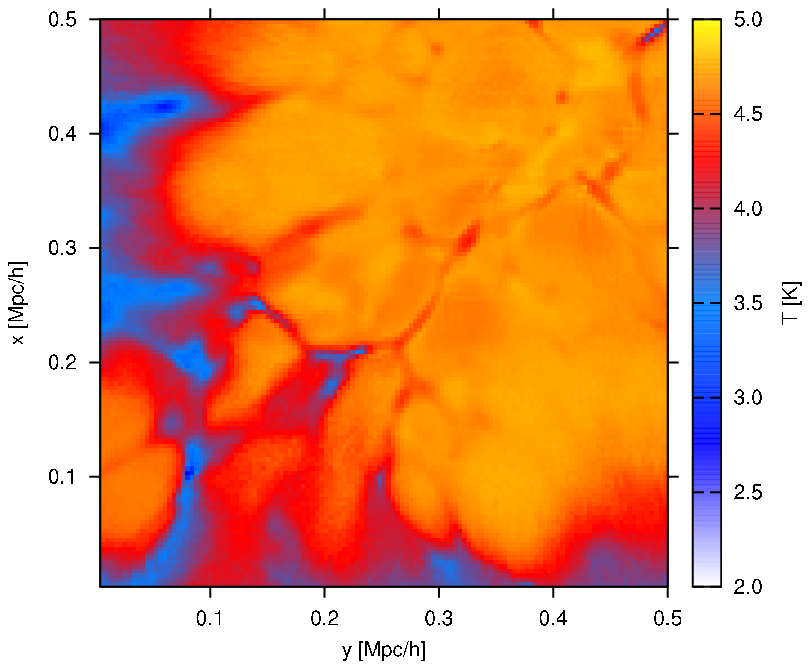}
 \hspace{5mm}
 \includegraphics[width=84mm]{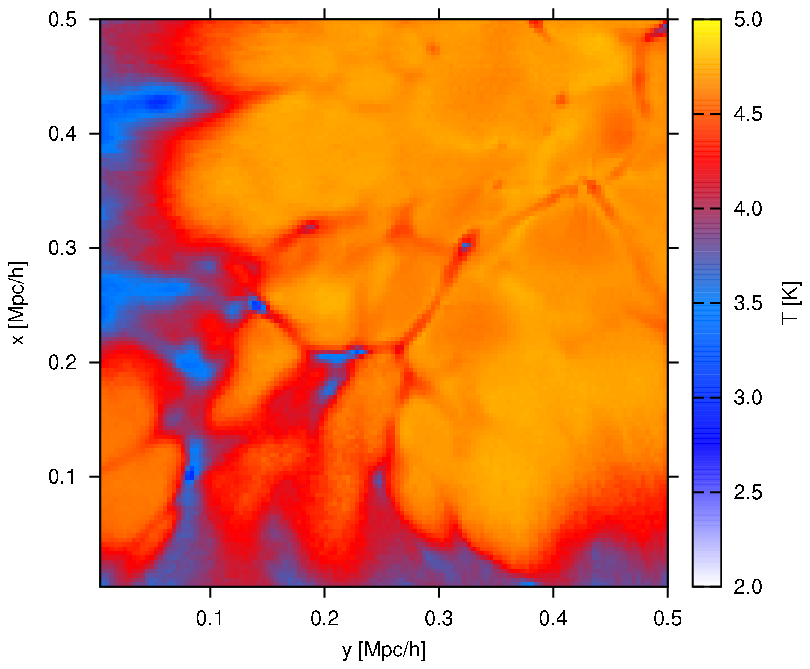} \\

 \includegraphics[width=84mm]{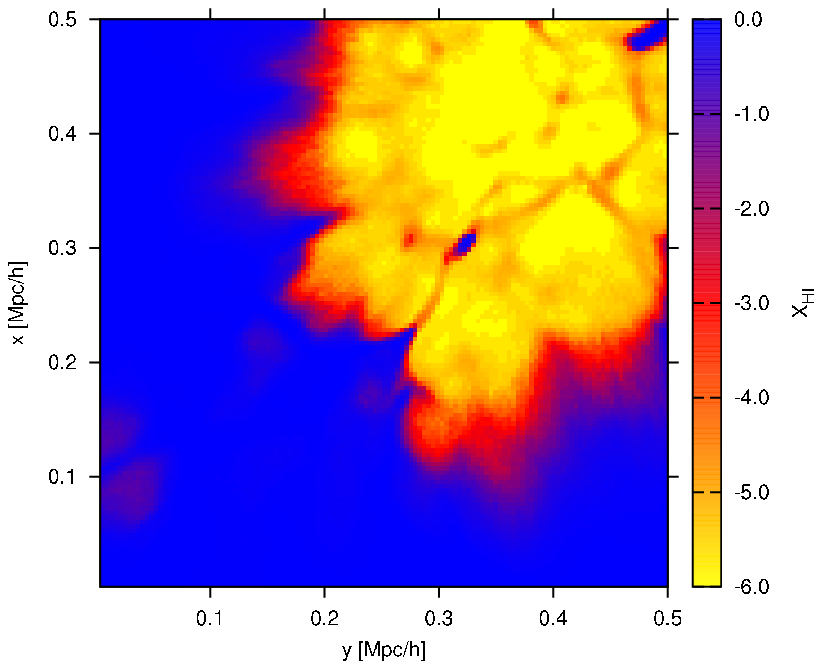}
 \hspace{5mm}
 \includegraphics[width=84mm]{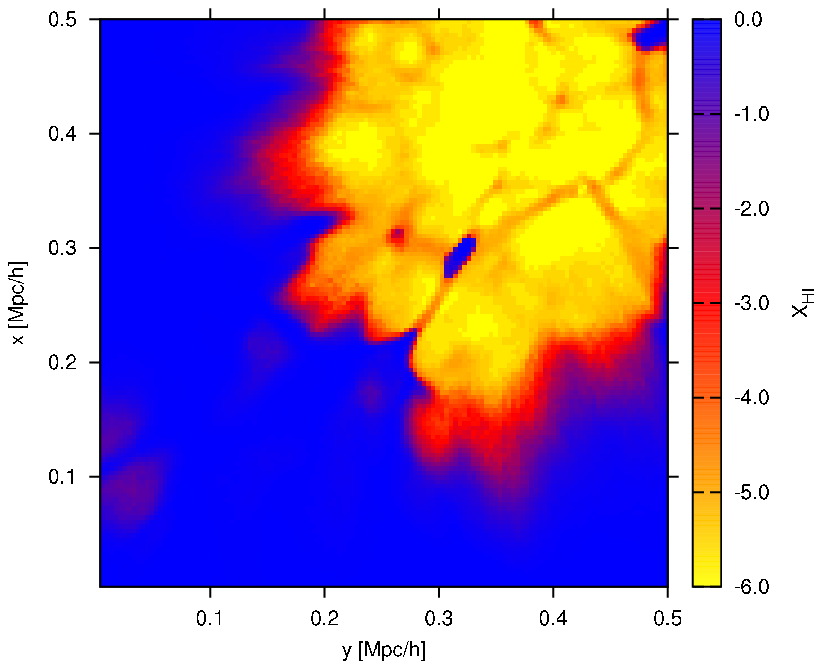} \\

 \includegraphics[width=84mm]{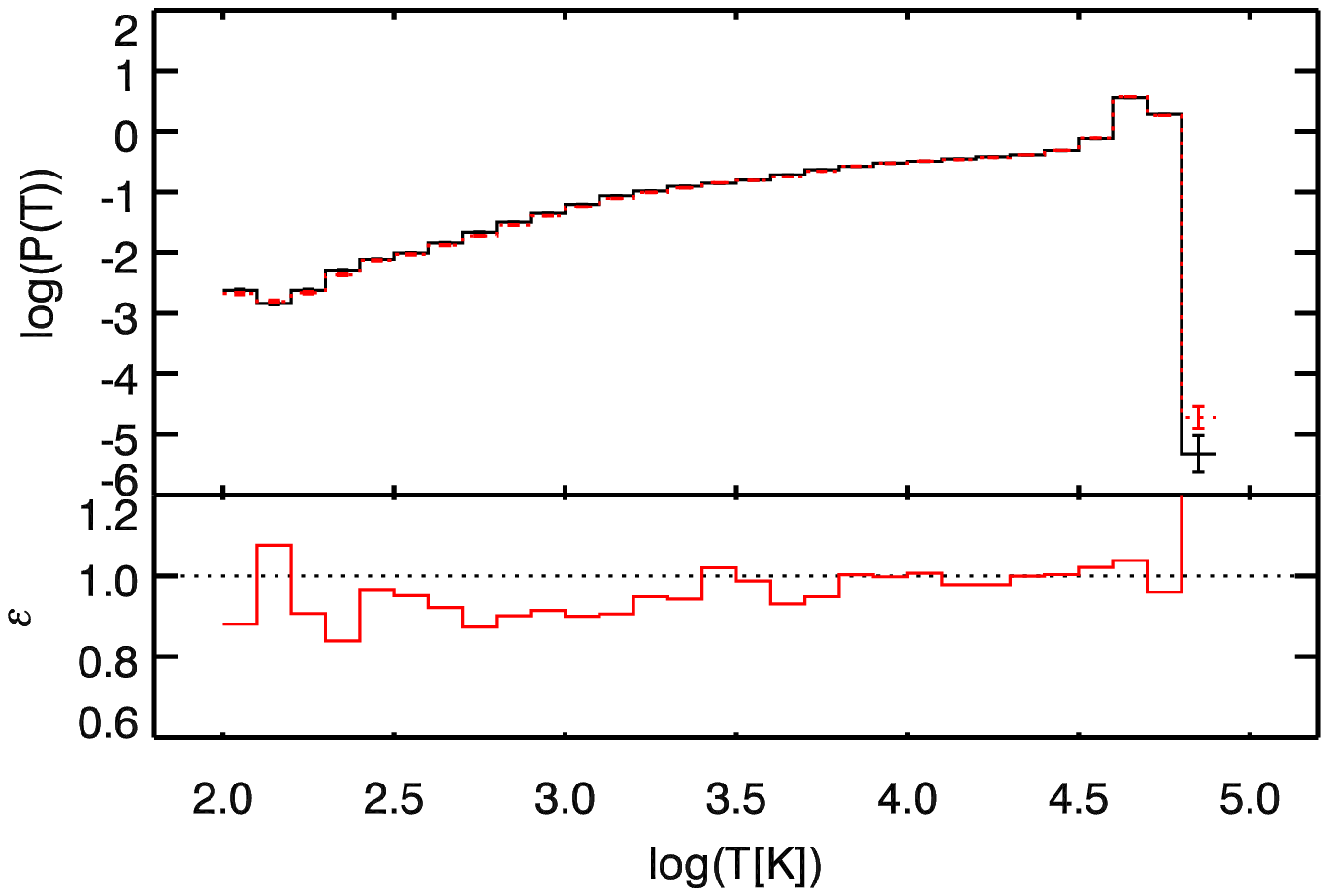}
 \hspace{5mm}
 \includegraphics[width=84mm]{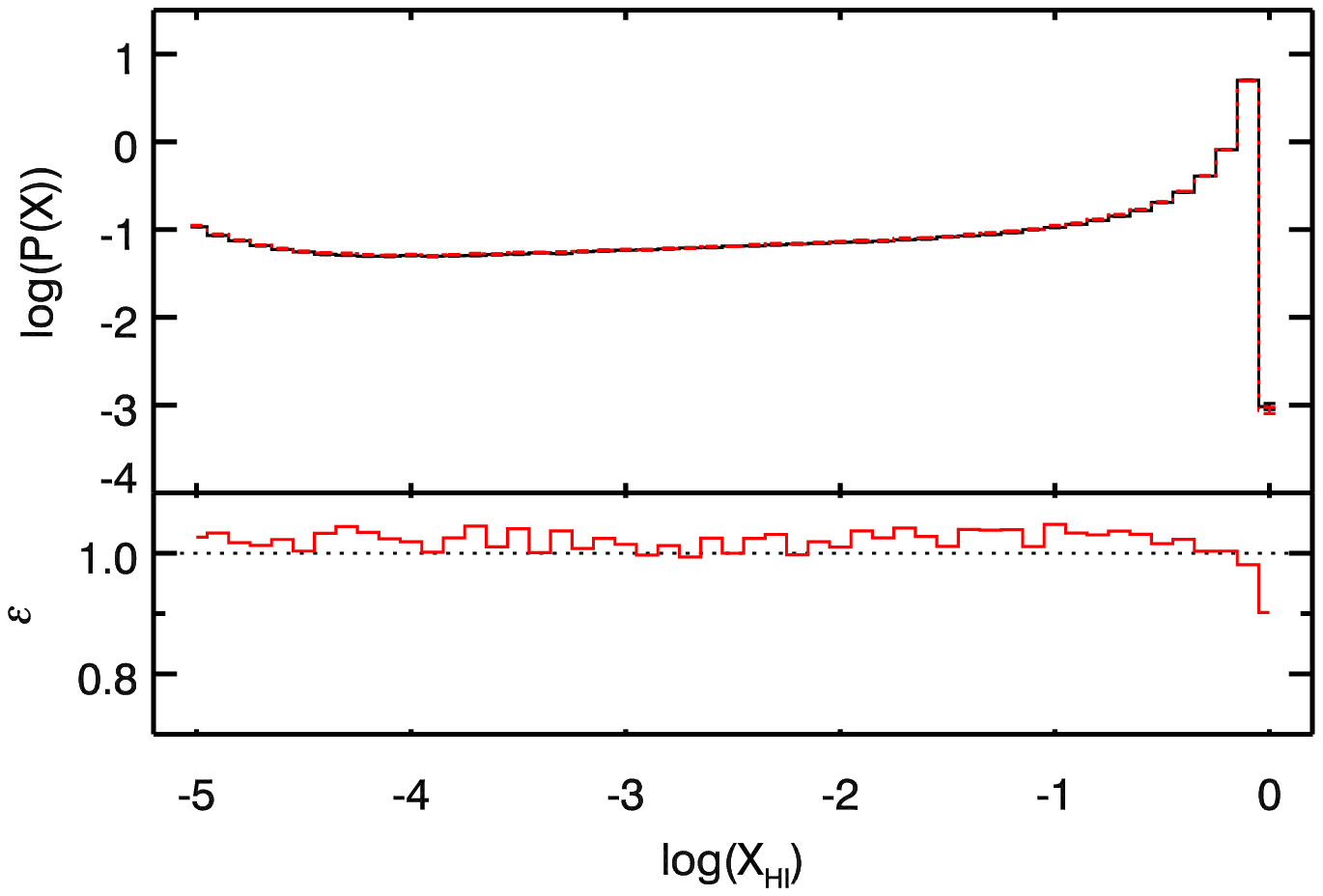} \\

 \caption{Test 4: Same as in Fig. \ref{fig:test4_time-step1} at time-step $t=0.2 \textrm{ Myr}$.}
 \label{fig:test4_time-step2}
\end{figure*}

A comparison between slices of the neutral hydrogen fraction and the temperatures obtained
with {\tt pCRASH2} and {\tt CRASH2} are given in Fig. \ref{fig:test4_time-step1} for 
$t=0.05 \textrm{ Myr}$ and Fig. \ref{fig:test4_time-step2} for $t=0.2 \textrm{ Myr}$. For both time 
steps the {\tt pCRASH2} results produce qualitatively similar structures compared to {\tt CRASH2}
in the neutral fraction and 
temperature fields. Slight differences however are present, mainly in the vicinity of the
ionisation fronts, due to the differences in how 
recombination is treated, as already discussed. In the lower panels of the same figures we show also 
the probability distribution functions for the hydrogen neutral fractions and the temperatures. 
Here the differences are more evident. The neutral fractions obtained with {\tt pCRASH2} do not show 
strong deviations from the {\tt CRASH2} solution.
In the distribution of temperatures however, {\tt pCRASH2} tends to heat up the initially cold
regions somewhat faster than {\tt CRASH2}, as can be seen by the 20\% drop in probability for the
$t=0.05 \textrm{ Myr}$ time step at low temperatures. This can be explained by the fact that the dominant component to the ionising radiation field
in the outer parts of \ion{H}{ii} regions is given by the radiation emitted in recombinations. 
The better resolution adopted in 
{\tt pCRASH2} for the diffuse field is then responsible for slightly larger \ion{H}{ii} regions. Since
a larger volume is ionised and thus hot, the fraction of cold gas is smaller in the {\tt pCRASH2}
run, than in the reference run. At higher temperatures however, the
distribution functions match each other well. Only in the highest bin a discrepancy between the
two code's solutions can be noticed, but given that there are at most four cells contributing to that bin, this is consistent with Poissonian
fluctuations. 

\subsection{Scaling properties}
\label{sec:scaling}

By studying the weak and strong scaling properties of a code, it is possible to assess
how well a problem maps to a distributed computing environment and how much speed
increase is to be expected from the parallelisation. 
Strong scaling describes the scalability by keeping the 
overall problem size fixed. Weak scaling on the other hand refers to how the scalability 
behaves when only the problem size per core is kept constant (i.e. the fraction between 
the total problem size and the number of CPUs stays constant). 

We will use the test cases described above, to study
the impact of the various input parameters such as grid size, number of photon packets, number
of sources, and type of physics used, on the scalability. Scalability is measured with the 
speed up, which is defined as the execution time on a single core divided by the execution
time on multiple cores. Due to the long simulation times of our test cases on a single core
(for some test cases of the order of weeks), the execution time on a single core 
was only determined once. This certainly makes our normalisation point subject to 
variance, which has to be taken into account in the following discussion.
All results presented in this section have been obtained on
a cluster at the Astrophysikalisches Institut Potsdam (AIP)
with computational nodes consisting of two Intel Xeon Quad Core 2.33GHz processors each. Since our parallelisation scheme has shortly spaced communication 
points, very good communication latency is required, which on our system is guaranteed through 
an InfiniBand network.

\subsubsection{Test 2}
\label{sec:scale_test2}

\begin{figure}
 \includegraphics[width=84mm]{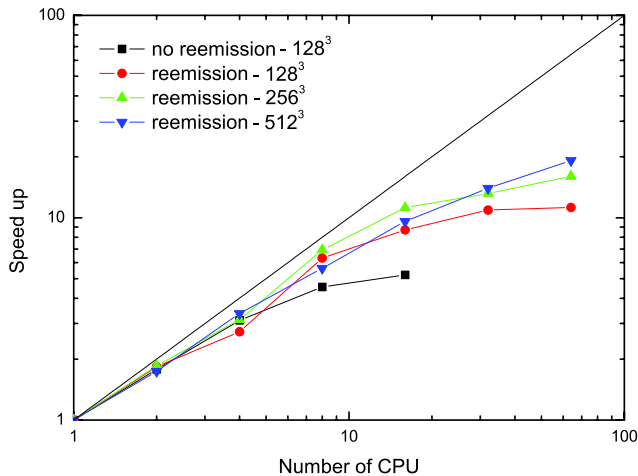}

 \caption{Speed up achieved on our local cluster for Test 2 as a function of grid-size and 
 number of CPUs. The straight black line indicates a linear ideal scaling property.}
 \label{fig:speedUp_test2}
\end{figure}

Achieving good scaling for Test 2 is a challenge, since only one source sits in a corner 
of the box and
a strong load imbalance between the sub-domains near to the source and the ones 
further away
exists. The fact that the gas is initially neutral even intensifies the imbalance, since at first
the domains far away from the source will be idle. Only as the \ion{H}{ii} region grows, more
and more sub-domains will be involved in the calculation.
This test is therefore a good proxy for how the code scales in extreme situations and shows
the strong scaling relation for one source. In Fig. 
\ref{fig:speedUp_test2} we show the results obtained with the setup described above. 
In principle it is expected, that the more rays each domain has to process in one time-step,
the better the scaling becomes. Therefore we run the test once without recombination emission
and once with. Further we increased the grid size to $256^3$ and $512^3$ cells, which increases
the amount of calculations needed per sub-domain. This yields the weak scaling properties 
for one source as a function of grid size.

The scaling relation is mainly determined by Ahmdal's law \citep{Amdahl:1967cs}, 
which relates the parallelised
parts of a code (in our case the propagation of photons) with the unparallelised parts (such
as communication between sub-domains). It states that the maximum possible speed up is
solely limited by the fraction of time spent in serial parts of the code.
The scaling relation behaves for all runs similarly. An almost linear speedup is achieved
up to 8 CPUs per node, and a deviation from 
linear scaling according to Ahmdal's law arises when communication over the network is needed. 
The deviation is dependent on the amount of work to be processed per time-step. 

In the worst case of a $128^3$ celled grid without reemission, the deviation is the largest. 
Following recombination emission and thus increasing the workload yields better scaling
up to the point where one computational node is saturated. For one source the best scaling 
is achieved with larger grid sizes. For the $512^3$ grid it is even reasonable to 
use 16 CPUs on two nodes on our cluster  and allow for communication of rays over the 
network. However with 32 CPUs, 50\% of the computational resources remain unused.

\begin{figure}
 \includegraphics[width=84mm]{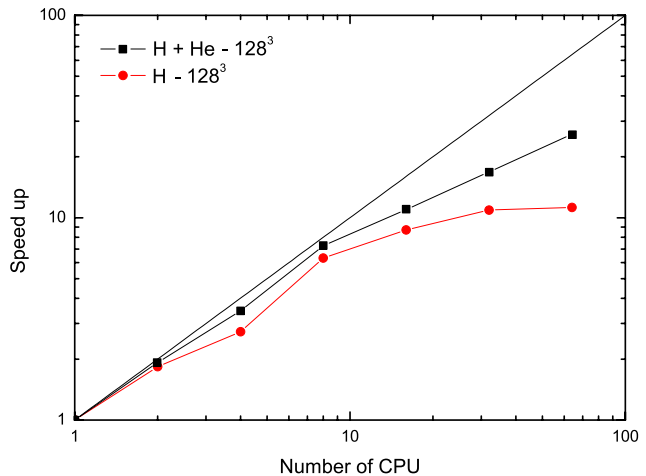}

 \caption{Test3: Scaling relation of the Cloudy test case as a function of number of CPUs.
For comparison, the
 results of the hydrogen only Test 2 are given as red line.}
 \label{fig:speedUp_testCC}
\end{figure}

\subsubsection{Test 3}
\label{sec:scale_testCC}

Up to now, we have only considered the gas to be made up solely of hydrogen. We now study
the scaling properties of one source embedded in gas made of a mixture of hydrogen and helium. 
Now each domain needs
to solve a more complicated set of equations and more time is spent in solving the chemical-thermal equations network  than in propagating photons.
The ionisation network needs to track the two additional species introduced with helium, plus the related contribution to heating and cooling which enters the 
temperature calculation. 
Since solving the ionisation-thermal network is the most computationally expensive part of our algorithm, the additional
work results in an approximate increase of overall execution time by a factor of three or more. Furthermore 
recombining helium will produce additional photons that need to be followed, even further 
increasing the computational load.
As the communication demand stays the same as in Test 2, the scaling relation is expected to 
be better than in the hydrogen only case.
In Fig. \ref{fig:speedUp_testCC} we present the scaling performance obtained with the Cloudy
test case described above using $N_{\rm{p}} = 2 \times 10^9$ photon packets and
$N_{\rm{t}} = 10^7$ time-steps. 
Comparing the scaling with the one achieved in Test 2, it is evident, that solving more detailed physics 
improves scaling. Now for the Cloudy test, ideal scaling is achieved as long as the problem is
confined to one node (in our case 8 CPU) and no information needs to be communicated 
over the network. However
the fact that more time is spend in solving the rate equations, communication latency across
the network does not affect scaling as strongly as in Test 2. It now even makes sense to
use 32 CPUs for one source, yielding a speed up of around 16. As a comparison Test 2
achieved a speed up of around 10 for 32 CPUs.

As a final check we have also run a set of simulations for Test 3 with an increased gas density, $n_H=10$ cm$^{-3}$, all the other quantities being the same. The aim of this experiment is to test the scaling under conditions in which diffuse radiation from recombinations becomes important. Somewhat to our surprise, we do not find any appreciable differences in the scaling relation between the low- and high-density cases. However, this can be easily understood as follows: The simulation time is set at five times the recombination time. Hence the same amount of recombinations per time step occur (when normalised to the density of the gas) in both runs. Since recombination photons are produced when a certain fraction of the gas has recombined, the number of emitted recombination photons is similar in the two runs. Therefore the amount of CPU time spent in the diffuse component is similar and the scaling does not change. 

\subsubsection{Test 4}
\label{sec:scale_test4}

By studying the scaling properties of Test 4, we can infer how the code scales with increasing
number of sources. Since Test 4 uses an output of a cosmological simulation, the sources
are not distributed homogeneously in the domain. Therefore large portions of the grid 
remain neutral as is seen in Fig. \ref{fig:test4_time-step2}. This poses a challenge to our 
domain decomposition strategy and might deteriorate scalability. 

\begin{figure}
 \includegraphics[width=84mm]{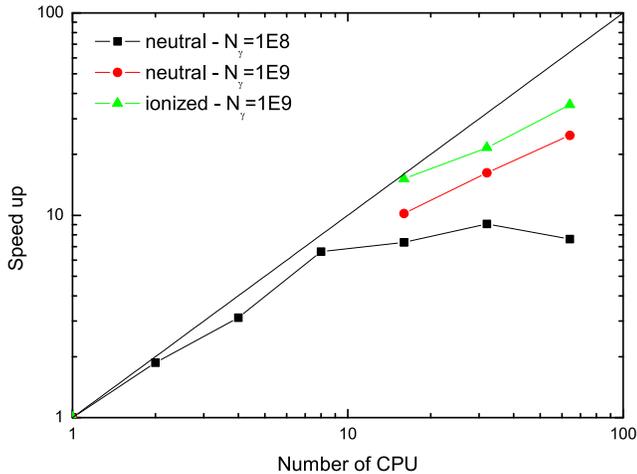}

 \caption{Scaling achieved with Test 4 as a function of number of photon, 
 optically thickness and number of CPUs.}
 \label{fig:speedUp_test4_numPhot}
\end{figure}

First we study the scaling of the original Test 4, with a sample of $10^8$ photon packets 
emitted
by each of the 16 sources. Then we increase the number of photons per source to $10^9$.
Since large volumes remain neutral, we further study an idealised case, where the whole box
is kept highly ionised by initialising every cell to be 99\% ionised. 
The results of these experiments are shown in Fig. 
\ref{fig:speedUp_test4_numPhot}.
In the case of only $10^8$ photon packets per source, good scaling can only be reached up
to 8 CPUs (i.e. a single node), as was the case with only one source. Communication 
over the network 
cannot be saturated with such a small number of photons and its overhead is larger than the 
time spent in computations. However by increasing the number of samples per time-step 
improves scaling. In the idealised optically thin case, perfect
scaling is reached even up to 16 CPUs and starts to degrade for higher numbers of CPU.
The discrepancy between the neutral and fully ionised case shows the influence of load
imbalance due to the large volumes of remaining neutral gas.

\begin{figure}
 \includegraphics[width=84mm]{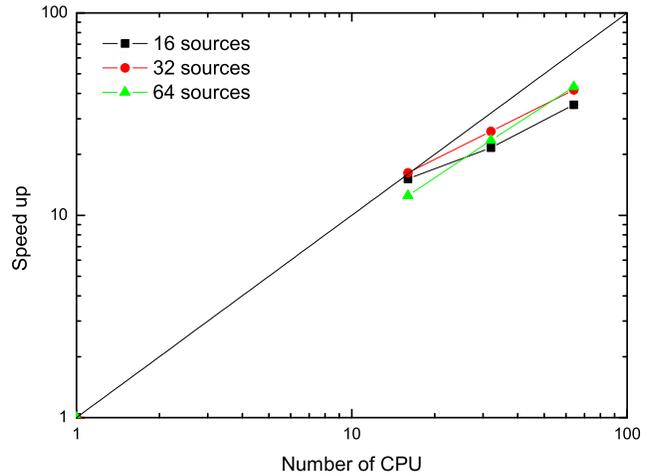}

 \caption{Scaling achieved with Test 4 as a function of number of sources and number of
 CPUs for an ideal optically thin case.}
 \label{fig:speedUp_test4_numSources}
\end{figure}

Since good scaling was found in the original Test 4, we now increase the number
of sources. For this we duplicate the sources in Test 4, and mirror their position at an arbitrary
axis. This preserves the clustered nature of the sources, while distributing them throughout the computational volume. 
With this prescription we increase the number of sources to 32 and 64. Again each source emits $10^9$ photon packets and
the idealised case of an optically thin box is studied. The results of the speed up function are 
given in Fig. \ref{fig:speedUp_test4_numSources}. As expected, the more computational 
intensive the problem becomes, the better the scaling. In the figure, the speed up for the 64
sources lies below the others cases because we have linearly extrapolated the
normalisation point of the 32 sources run to the 64 sources case, since the time to run the
simulation on one CPU was too long. 
This of course is just a rough estimate, and in fact it is responsible
for the lower speed up values. Despite of it, the trend still gives a measure of the improved scaling with respect to the cases with 32 and 16 sources. 
While a difference in the speed up between the three test-cases is seen when increasing from 1 to 16 CPUs, it 
is difficult to tell whether this is a genuine feature of our parallelisation strategy or just a 
manifestation of the variance in the normalisation.

\begin{figure}
 \includegraphics[width=84mm]{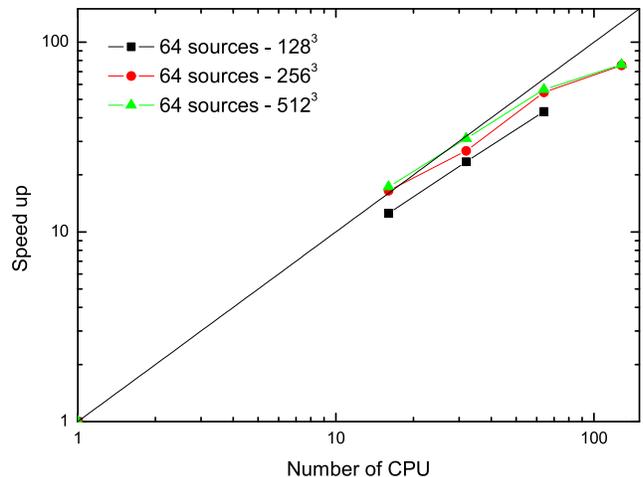}

 \caption{Scaling relation of Test 4 as a function of box size with 64 sources and number of 
 CPUs for an ideal optically thin case.}
 \label{fig:speedUp_test4_boxSize}
\end{figure}

Up to now we have only considered the case of a $128^3$ box. We now re-map the
density field of Test 4 to $256^3$ and $512^3$ grids. The resulting scaling properties for the 
ideal optically thin case with 64 sources are presented in Fig. \ref{fig:speedUp_test4_boxSize}.
The scaling properties are not dependent on the size of the grid, and weak scaling only
exists as long as the number of cores does not exceed the number of sources. 

We can thus conclude that our parallelisation strategy shows perfect weak scaling 
properties when increasing the numbers of sources (compare Fig. 
\ref{fig:speedUp_test4_numSources}). Increasing the size of the grid does not 
affect scaling. However weak scaling in terms of the grid size only works as long as the 
number of CPUs does not exceed the number of sources (see Fig. 
\ref{fig:speedUp_test4_boxSize} going from 64 to 128 CPUs).

From these tests, we can formulate an optimal choice for the simulation setup. We have seen
that better scaling is achieved with large grids. 
Further linear scaling can be achieved in the optically thin case by using up to as 
may cores as there are sources. For the optically thick case however, deterioration of 
the scaling properties needs to be taken into account. 

\begin{figure*}
 \includegraphics[width=84mm]{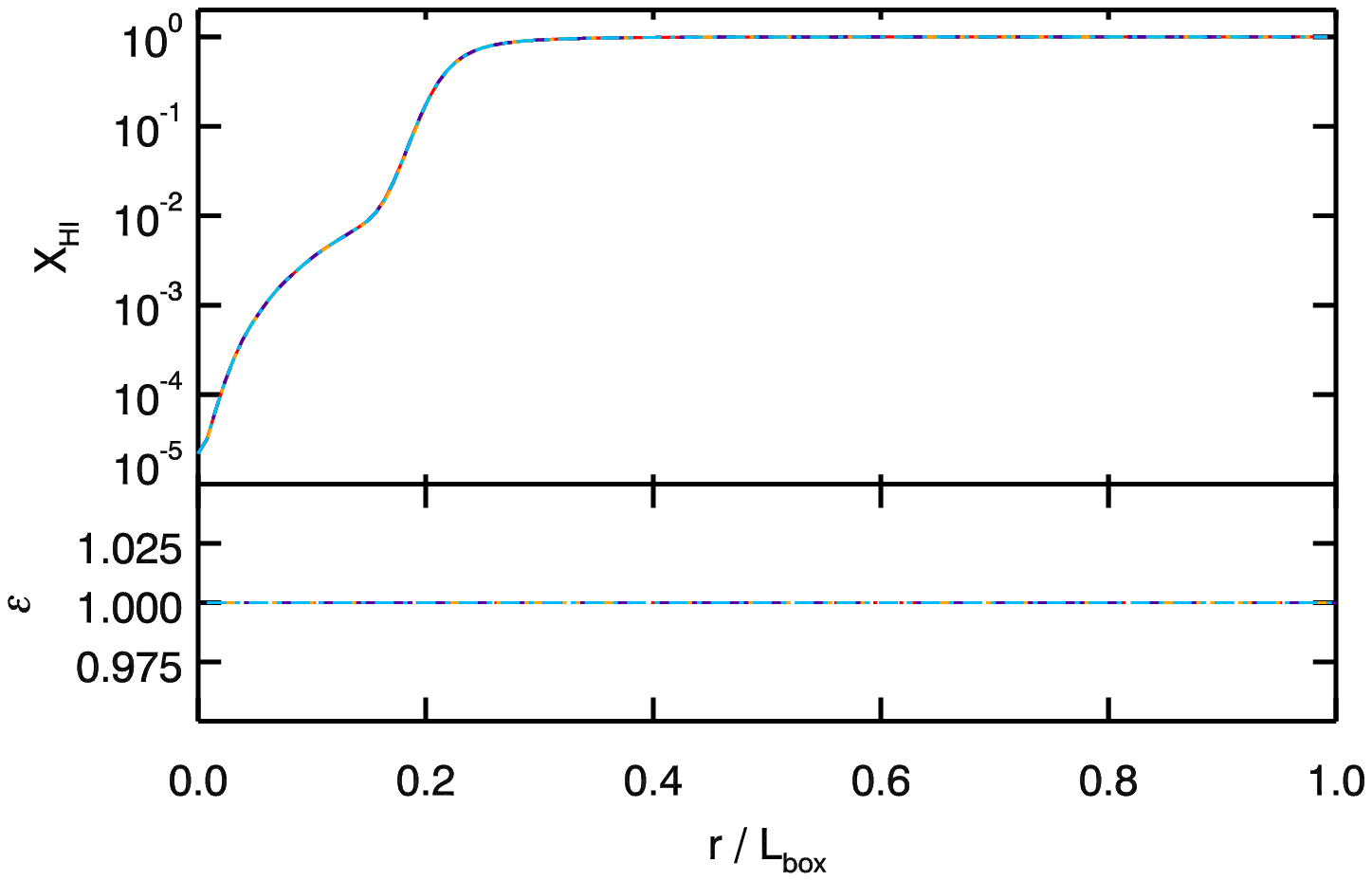}
 \hspace{5mm}
 \includegraphics[width=84mm]{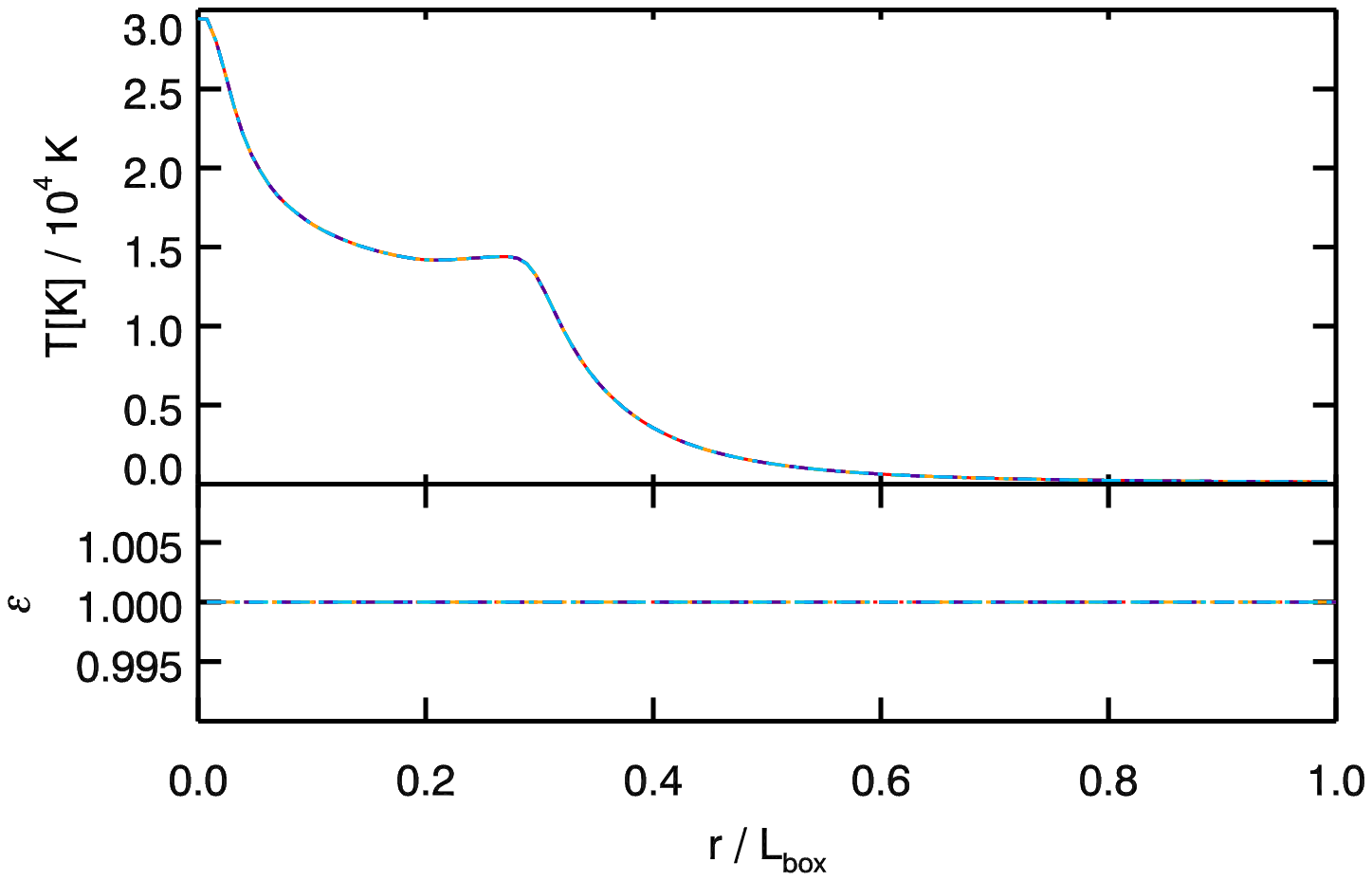} \\

 \includegraphics[width=84mm]{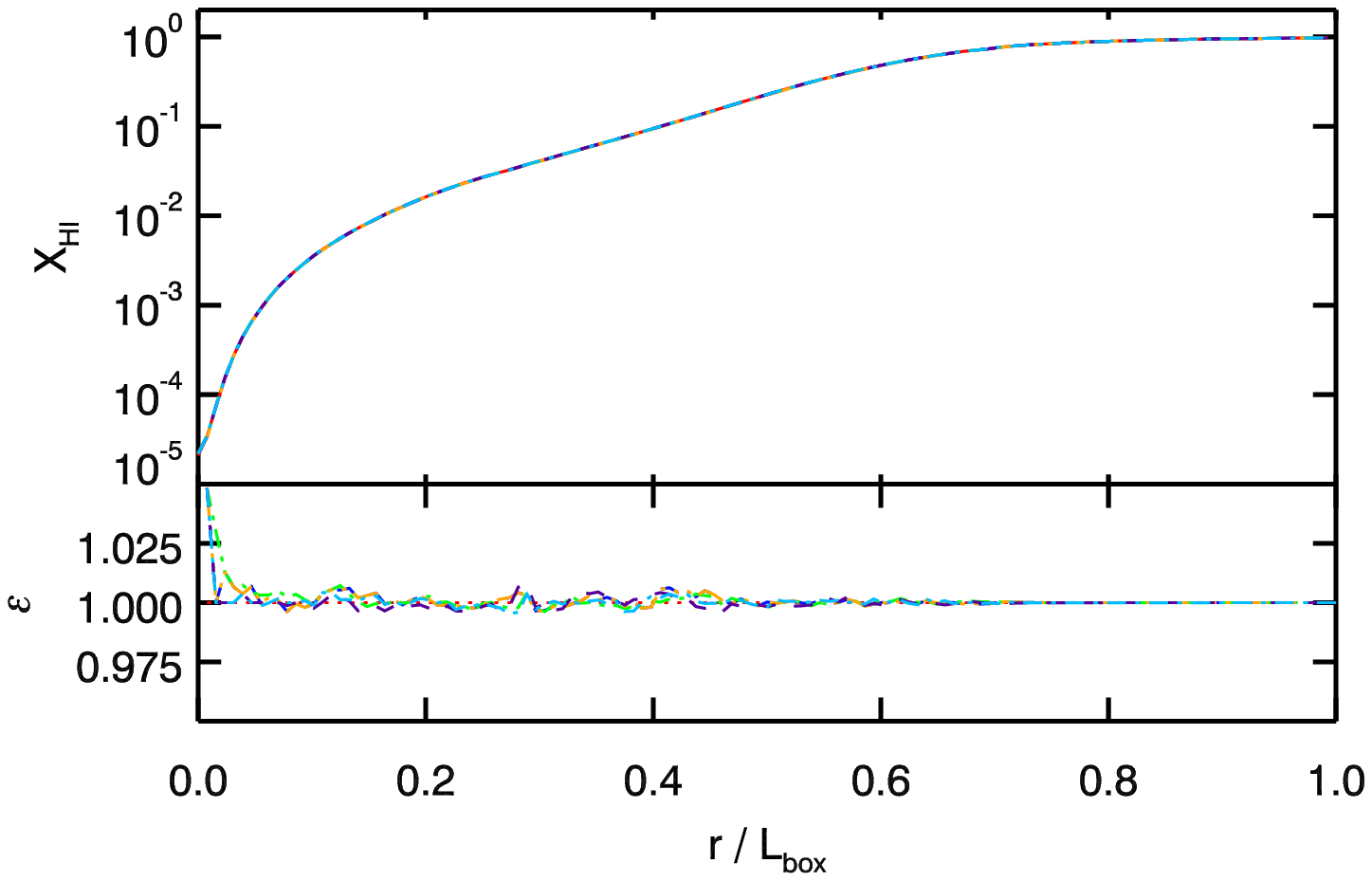}
 \hspace{5mm}
 \includegraphics[width=84mm]{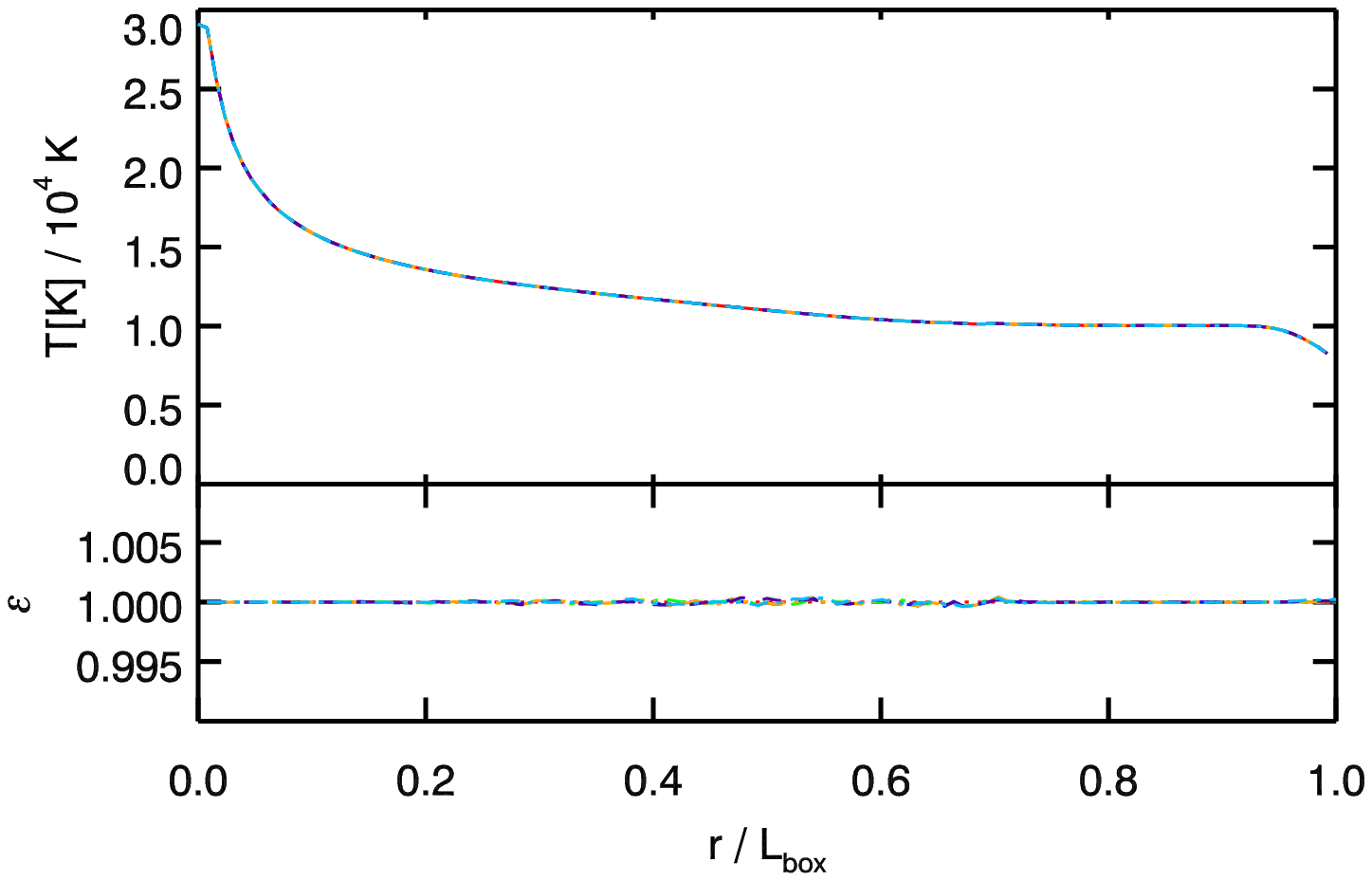} \\

 \caption{Test 2: Dependence of the solution on the number of cores. The large upper panels 
 show the results of {\tt pCRASH2} obtained on 1 CPU (red dotted line), 2 CPUs (blue dashed
 line), 4 CPUs (green dash dotted line), 8 CPUs (orange dash dot dotted line), 16 CPUs (violet
 dashed line), and 32 CPUs (light blue dash dotted line). The results are compared with the
 {\tt CRASH2} solution (black solid line). 
 Shown are the spherically averaged neutral hydrogen fraction profiles of the sphere 
 as a function of radius (left column) and the temperature profiles (right column) at 
 $t=1\times10^7$ (top), and $5\times10^8 \textrm{ yr}$ (bottom). 
 The small bottom panels show the relative deviation from the single CPU {\tt pCRASH2}
 solution.}
 \label{fig:test2CompCPUs}
\end{figure*}

\subsection{Dependence of the solution on the number of cores}
\label{sec:cpuDependance}

Since each core has its own set of random numbers, the solutions of the same problem
obtained with different numbers of cores will not be identical. They will vary according to
the variance introduced by the Monte Carlo sampling. To illustrate the effect, we
revisit the results of Test 2 and study how the number of cores used to solve the problem
affects the solution.

The results of this experiment are shown in Fig. \ref{fig:test2CompCPUs}, where we compare
the different solutions obtained with various numbers of CPUs. Solely by looking at the profiles,
no obvious difference between the runs can be seen. Variations can only be seen
in the relative differences of the various runs which are compared with the single CPU 
{\tt pCRASH2} run. 

At $t=10^7 \textrm{ yr}$ the runs do not show any differences. In Sec. 
\ref{sec:test2} we have seen, that recombination radiation has not yet started to be important. 
This is exactly
the reason why, at this stage of the simulation, no variance has developed between the different runs.
Since the source is always handled by the first core and the set of random number
is always the same on this core no matter how many CPUs are used, the results are always
identical. However as soon as the Monte Carlo sampling process start to occur on multiple nodes, 
the set of random numbers starts to deviate from the single CPU run and variance in the sampling is introduced.
At the end of the simulation at $t=5\times10^8 \textrm{ yr}$, large parts of the \ion{H}{ii} region
are affected by the diffuse recombination field and variance between the different runs
is expected. By looking at Fig. \ref{fig:test2CompCPUs}, the Monte Carlo variance for the
neutral hydrogen fraction profile lies between 1\% and 2\%. The temperature profile is not as 
sensitive to variance as the neutral hydrogen fraction profile. For the temperature the variance
lies at around 0.1\%.

\begin{figure*}
 \includegraphics[width=84mm]{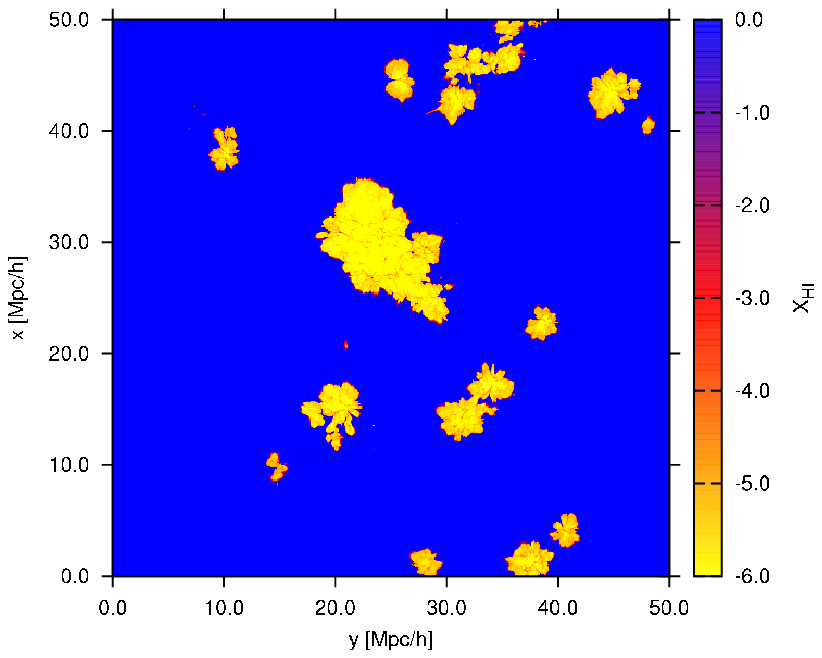}
 \hspace{5mm}
 \includegraphics[width=84mm]{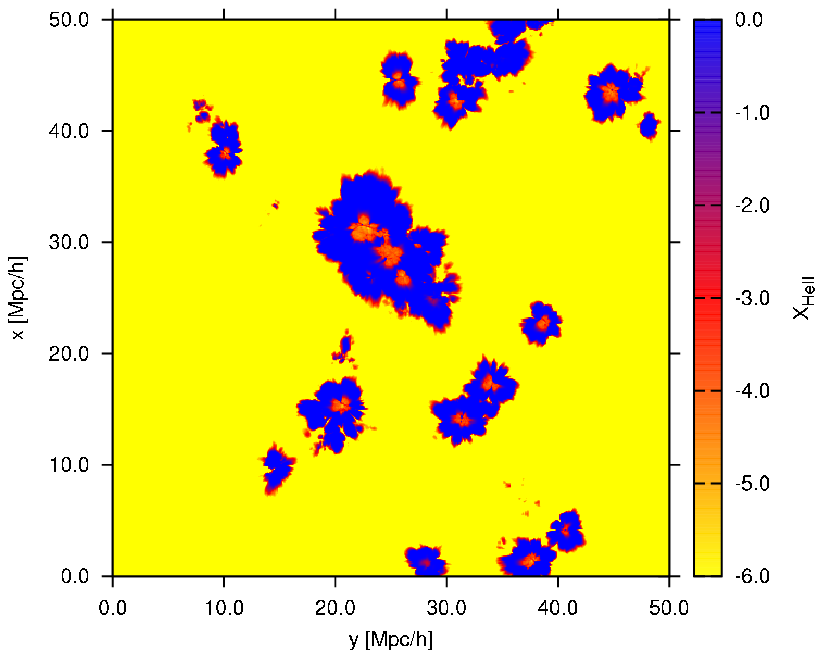} \\

 \includegraphics[width=84mm]{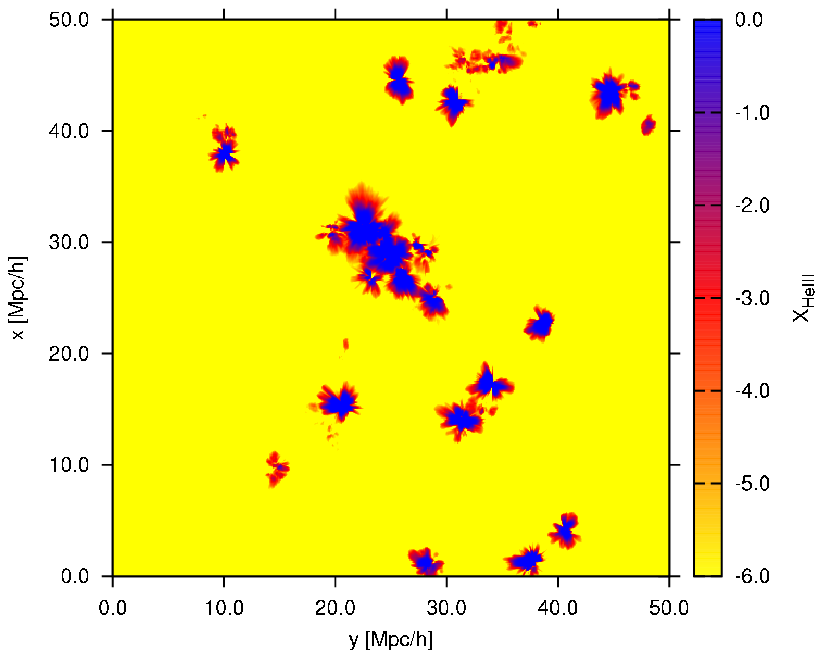}
 \hspace{5mm}
 \includegraphics[width=84mm]{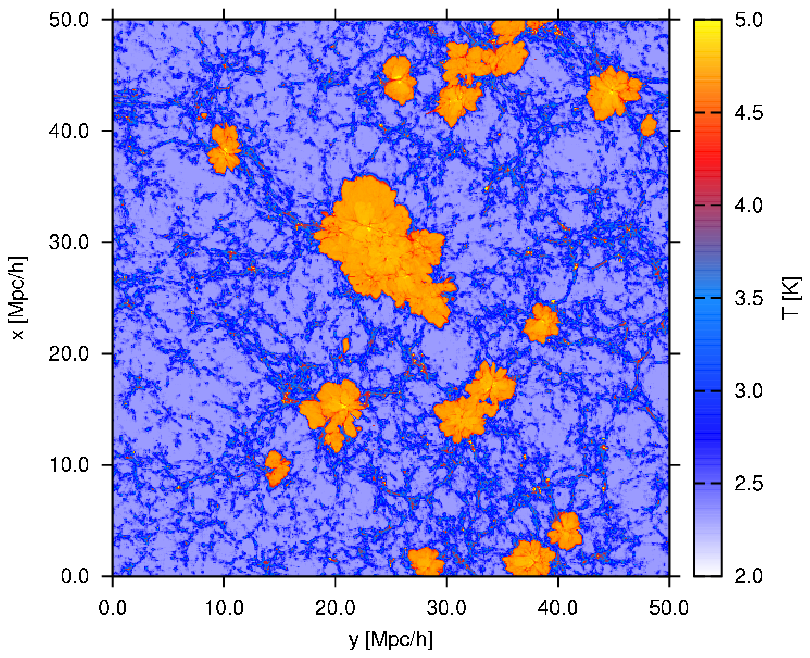} \\

 \caption{Cut through a cosmological simulation at $z=8.3$ with the neutral hydrogen fraction field 
 (upper left panel), the singly ionised helium
 fraction (upper right panel), the doubly ionised helium fraction (lower left panel), and the
 temperature field (lower right panel) at time-step $t=1 \times 10^6 \textrm{ yr}$ using
 1000 sources.}
 \label{fig:1000sources}
\end{figure*}

\subsection{Thousand sources in a large cosmological density field}
\label{sec:1000sources}

Up to now we have discussed {\tt pCRASH2}'s performance with controlled
test cases. We now demonstrate {\tt pCRASH2}'s ability to handle large highly
resolved cosmological density fields embedding thousands of sources. We 
utilise the $z=8.3$ output of the \emph{MareNostrum High-z Universe} \citep{Forero-Romero:2010mq}
which is a $50 \textrm{ Mpc } h^{-1}$ SPH simulation using $2\times1024^3$ particles
equally divided into dark matter and gas particles. The gas density and internal energy
are assigned to a $512^3$ grid using the SPH smoothing kernel. A hydrogen mass fraction
of $76 \%$ and a helium mass fraction of $24 \%$ are assumed.

The UV emitting sources are determined similarly to the procedure used in Test 4 
of the comparison project 
\citep{Iliev:2006fk} by using the 1000 most massive haloes in the simulation. We 
evolve the radiation transport simulation for $10^{6} \textrm{ yr}$ and follow
the hydrogen and helium ionisation, as well as photo-heating. Recombination emission is 
included. Each source emits $10^8$ photon packets in $10^7$ time 
steps. In total $10^{11}$ photon packets are evaluated, not counting 
recombination events. The simulation was run on a cluster using 16 nodes,
with a total of 128 cores. The walltime for the run as reported by the queuing system
was 143.5 hours; the serial version would have needed over two years to finish
this simulation.

In Fig. \ref{fig:1000sources} cuts through the resulting ionisation fraction fields and the
temperature field are shown at $t=10^6 \textrm{ yr}$. It can be clearly seen 
that the \ion{H}{ii} regions produced by the different sources already overlap each other.
Further the ionised regions strongly deviate from spherical symmetry. This is caused by
the fact that the ionisation fronts propagate faster in underdense regions than in dense
filaments. 

The distribution of \ion{He}{ii} follows the distribution of ionised hydrogen, 
except in the centre of the ionisation regions, where helium becomes doubly ionised and
holes start to emerge in the \ion{He}{ii} maps. Photo-heating increases the temperature
in the ionised regions to temperatures of around $30000 \textrm{ K}$.

These results are presented here as an example of highly interesting problems in cosmological studies which can be easily addressed with 
{\tt pCRASH2}, but  that would be impossible to handle with the serial version of the code {\tt CRASH2}.

\section{Summary}
\label{sec:conclusions}

We have developed and presented {\tt pCRASH2}, a new parallel version of the {\tt CRASH2} 
radiative transfer scheme code, whose description can be found in \citet{Maselli:2003}, \citet{Maselli:2009gd} and references therein. The parallelisation strategy was developed to map the {\tt CRASH2} algorithm to distributed memory machines, using the MPI library.

In order to obtain an evenly load balanced parallel algorithm, we statically estimate the computational load in each cell by calculating the expected ray number density assuming an optically thin medium. The ray density in a cell is then inversely proportional to the distance to the source squared. Using the Peano-Hilbert space filling curve, the domain is cut into sub-domains; the integrated ray number density in the box is determined and equally divided by the number of processors. Then the expected ray number density is integrated along the curve and cut, whenever this fraction of the total ray number density is reached. The result of this is a well balanced domain decomposition, that even performs adequately in  optically thick regimes.

For the parallelisation of the ray tracing itself, we have segmented the propagation of rays over multiple
time steps. In the original {\tt CRASH2} implementation it was assumed that photons instantly propagate through the whole box in one time step. In 
{\tt pCRASH2} however rays are only propagated through one sub-domain per time step. After
rays have propagated to the border of the sub-domain, they are then passed on to the neighbouring
domain for further processing in the following time step. With this segmentation strategy we keep 
communication between the domains low and locally confined. 

Since {\tt pCRASH2} is able to handle a larger
amount of photon packets than the serial version, we have additionally increased the sampling resolution of the diffuse recombination radiation. With this parallelisation strategy, we obtain a high-performing algorithm, that scales well for large problem sizes. We have extensively tested {\tt pCRASH2} against a standardised set of test cases to validate the parallelisation scheme. 

With {\tt pCRASH2} it is now possible to address a number of problems that require 
the tracking of the UV field produced by a large number of sources, that so far were not attainable for {\tt CRASH2} because of the high computational cost. 
A natural application is the evolution of the UV flux field, especially during the era of reionisation.  The phase transition of the initially neutral H+He intergalactic to its  fully ionised state can now be accurately followed in the large volumes required to achieve a good representation of the process on cosmological scales 
($> 300 \textrm{Mpc}$).
  
On the other hand, {\tt pCRASH2} has a broader application range. In fact, it is not limited to tackle cosmological configurations, and can  be applied to study  a wealth of astrophysical problems, e.g. the propagation and impact of UV flux field in the vicinity of star forming galaxies, to mention one example. Finally, due to {\tt pCRASH2}'s distributed memory approach, new physics such as extending the ionisation network to species other than hydrogen and helium can be now easily implemented. 

\appendix
\section{Generating the Peano-Hilbert curve}
\label{app:Hilbert}

Space filling curves present an easy method to systematically reach every point
in space and are usually fractal in nature. Various space filling curves such as the 
Peano curve, the Z-order, or the Peano-Hilbert curve exist. The most optimal in 
terms of clustering (i.e. that all points on the curve are spatially near to each other) 
is the Peano-Hilbert curve. We will now sketch the fast Peano-Hilbert algorithm used 
by the domain decomposition algorithm for projecting cells on the grid which is a
subset of the natural numbers including zero $\mathbb{N}_0^3$ onto an array in 
$\mathbb{N}_0^1$. A detailed description of its implementation is found in
\cite{Chenyang:2008vl}. 

Let $H_m^N$, $(m \ge 1, N \ge 2)$ describe an $N$-dimensional Peano-Hilbert
curve in its $m$th-generation. $H_m^N$ thus maps $\mathbb{N}_0^N$ to $\mathbb{N}_0^1$,
where we call the mapped value in $\mathbb{N}_0^1$ a Hilbert-key. A $m$th-generation
Peano-Hilbert curve of $N$-dimension is a curve that passes through a hypercube
of $2^m \times \ldots \times 2^m = 2^{mN}$ in $\mathbb{N}_0^N$. For our purpose
we only consider $N \le 3$.

\begin{figure}
 \includegraphics[width=84mm]{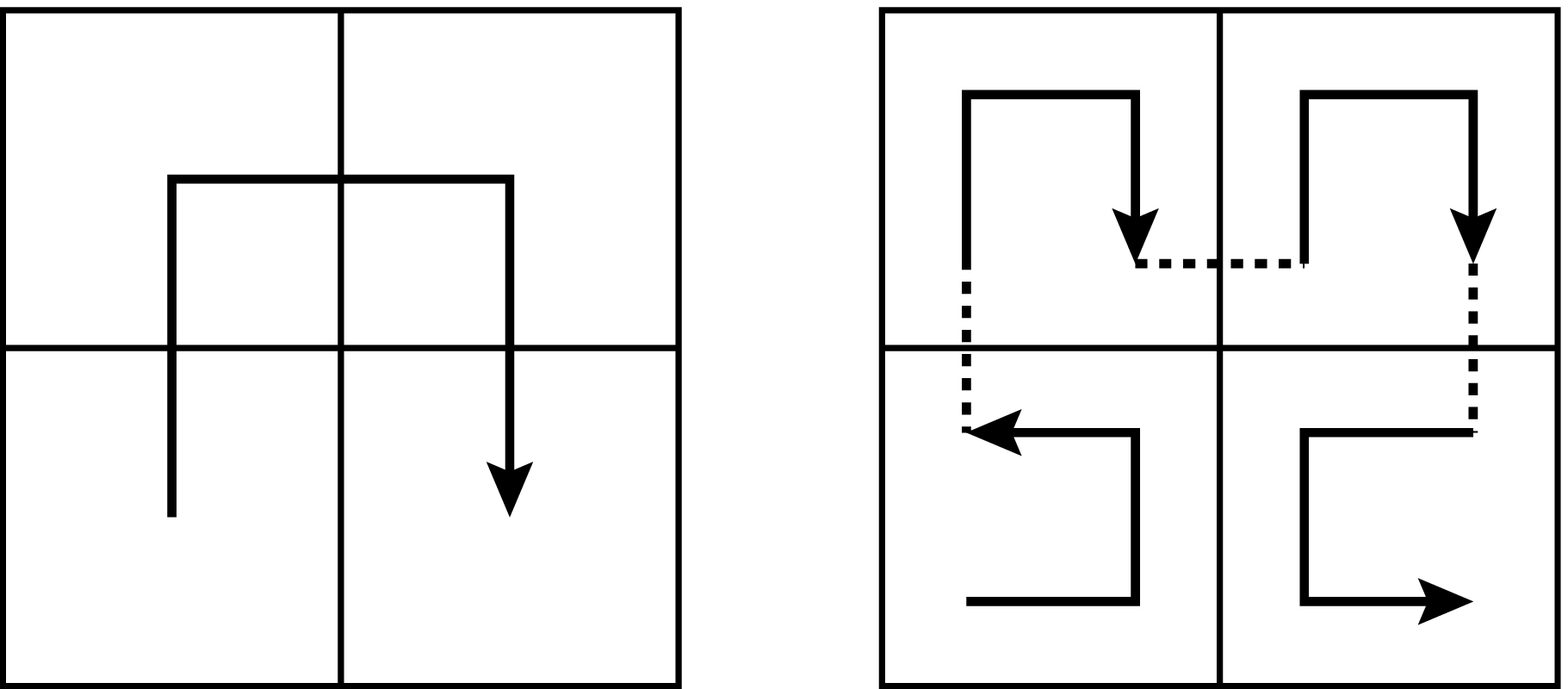}
 
 \caption{Left panel: The Hilbert cell $C^2 \equiv H_1^2$ for $N=2$. Right panel:
 Applying the $N=2$ Hilbert genes on the Hilbert cell produces the 2nd-generation
 Peano-Hilbert curve.}
 \label{fig:hilbert}
\end{figure}

Let the 1st-generation Peano-Hilbert curve be called a $N$-dimensional Hilbert cell $C^N$ 
(see Fig. \ref{fig:hilbert} for $N=2$). In binary digits, the coordinates of the 
$C^2[\textrm{Hilbert-key}]$ Hilbert cell can be expressed as $C^2[0] = 00$, $C^2[1] = 01$,
$C^2[2] = 11$, and $C^2[3] = 10$, where each binary digit represents one coordinate $X_N$
in $\mathbb{N}_0^N$ with the least significant bit at the end, i.e. $C^2[i] = X_2X_1$.
For $N=3$ the Hilbert cell becomes $C^3[0] = 000$, $C^3[1] = 001$, $C^3[2] = 011$, $C^3[3] = 010$,
$C^3[4] = 110$, $C^3[5] = 111$, $C^3[6] = 101$, $C^3[7] = 100$.

The basic idea in constructing an algorithm that maps $\mathbb{N}_0^N$ onto the 
$m$th-generation Peano-Hilbert 
curve is the following. Starting from the basic Hilbert cell, a set of coordinate transformations which
we call Hilbert genes is applied $m$-times to the Hilbert cell and the final Hilbert-key is obtained. 
An illustration of this method is shown in Fig. \ref{fig:hilbert}. Here the method of extending
the 1st-generation 2-dimensional Peano-Hilbert curve to the 2nd-generation is shown. Analysing
the properties of the Peano-Hilbert curve reveals that two types of coordinate transformations operating
on the Hilbert cell are needed for its construction. The exchange of coordinates and the reverse 
operation. The exchange operation $X_2 \leftrightarrow X_3$ on $011$ would result in $101$.
The reverse operation denotes a bit-by-bit reverse (e.g. reverse $X_1, X_3$ on $010$ results in
$111$). Using these transformations and the fact, that the $m$th-generation Hilbert-key can be 
extended or reduced to an $(m+1)$- or $(m-1)$th-generation Hilbert-key by bit-shifting the key by 
$N$ bits up or down, the Hilbert-key can be constructed solely using binary operations.

For the $N=2$ case illustrated in Fig. \ref{fig:hilbert} the Hilbert genes $G^N[\textrm{Hilbert-key}]$
are the following: $G^2[0] =$ exchange $X_1$ and $X_2$, $G^2[1] =$ no transformation,
$G^2[2] =$ no transformation, and $G^2[3] =$ exchange $X_1$ and $X_2$ plus reverse
$X_1$ and $X_2$. Through repeated application of these transformations on the $m$th-generation
curve, the $(m+1)$th-generation can be found.

The Hilbert genes for $N=3$ are: $G^3[0] =$ exchange $X_1$ and $X_3$, $G^3[1] =$ exchange
$X_2$ and $X_3$, $G^3[2] =$ no transformation, $G^3[3] =$ exchange $X_1$ and $X_3$ plus
reverse $X_1$ and $X_3$, $G^3[4] =$ exchange $X_1$ and $X_3$, $G^3[5] =$ no transformation,
$G^3[6] =$ exchange $X_2$ and $X_3$ plus reverse $X_2$ and $X_3$, and $G^3[7] =$ exchange
$X_1$ and $X_3$ plus reverse $X_1$ and $X_3$.

\section*{Acknowledgments}
This work was carried out under the HPC-EUROPA++ project (project number: 211437), 
with the support of the European Community - Research Infrastructure Action of the FP7 
ÒCoordination and support actionÓ Programme (application 1218).
A.P. is grateful for the hospitality received during the stay at CINECA, especially by R. Brunino and C. Gheller. 
We thank P. Dayal for carefully reading the manuscript.
Further A.P. acknowledges support in parts by the German Ministry
for Education and Research (BMBF) under grant FKZ 05 AC7BAA. 
Finally, we thank the anonymous referee for important improvements to this work.

\bibliographystyle{mn2e}
\bibliography{literature}

\label{lastpage}

\end{document}